\documentclass[twocolumn, prl, superscriptaddress,notitlepage]{revtex4-2}
\usepackage{graphicx}
\usepackage{dcolumn}
\usepackage{bm}
\usepackage[usenames,dvipsnames]{color}
\usepackage[most]{tcolorbox}
\usepackage{mathtools}
\DeclarePairedDelimiter{\ceil}{\lceil}{\rceil}
\usepackage{multirow}
\usepackage{gensymb}
\usepackage[normalem]{ulem}
\usepackage{CJK}
\usepackage{comment}
\usepackage[colorlinks, linkcolor=blue,anchorcolor=blue,citecolor=blue,urlcolor=blue]{hyperref}
\usepackage{amssymb}
\usepackage{pifont}
\usepackage{physics}
\usepackage{natbib}
\usepackage{xcolor}

\usepackage{color,soul}

\begin{document}
\begin{CJK*}{UTF8}{}
\title{Hyperfine-enhanced gyroscope based on solid-state spins}

\author{Guoqing Wang \CJKfamily{gbsn}(王国庆)}\email[]{gq\_wang@mit.edu}
\thanks{These authors contributed equally.}
\affiliation{
   Department of Nuclear Science and Engineering, Massachusetts Institute of Technology, Cambridge, MA 02139, USA}
\affiliation{
   Research Laboratory of Electronics, Massachusetts Institute of Technology, Cambridge, MA 02139, USA}
\affiliation{Department of Physics, Massachusetts Institute of Technology, Cambridge, MA 02139, USA}

\author{Minh-Thi Nguyen}\email[]{minhthin@mit.edu}
\thanks{These authors contributed equally.}
\affiliation{
   Research Laboratory of Electronics, Massachusetts Institute of Technology, Cambridge, MA 02139, USA}
\affiliation{Department of Physics, Massachusetts Institute of Technology, Cambridge, MA 02139, USA}

\author{Paola Cappellaro}\email[]{pcappell@mit.edu}
\affiliation{
   Department of Nuclear Science and Engineering, Massachusetts Institute of Technology, Cambridge, MA 02139, USA}
\affiliation{
   Research Laboratory of Electronics, Massachusetts Institute of Technology, Cambridge, MA 02139, USA}
\affiliation{Department of Physics, Massachusetts Institute of Technology, Cambridge, MA 02139, USA}

\begin{abstract}
Solid-state platforms based on electro-nuclear spin systems are attractive candidates for rotation sensing due to their excellent sensitivity, stability, and compact size, compatible with industrial applications. Conventional spin-based gyroscopes measure the accumulated phase of a nuclear spin superposition state to extract the rotation rate and thus suffer from spin dephasing. Here, we propose a gyroscope protocol based on a two-spin system that includes a spin intrinsically tied to the host material, while the other spin is isolated. The rotation rate is then extracted by measuring the relative rotation angle between the two spins starting from their population states, robust against spin dephasing. In particular, the relative rotation rate between the two spins can be enhanced by their hyperfine coupling by more than an order of magnitude, further boosting the achievable sensitivity.  The ultimate sensitivity of the gyroscope is limited by the lifetime of the spin system and compatible with a broad dynamic range, even in the presence of magnetic noises or control errors due to initialization and qubit manipulations.  Our result enables precise measurement of slow rotations and exploration of fundamental physics.  
\end{abstract}

\maketitle

\end{CJK*}	

\textit{Introduction.---} Inertial sensing finds broad applications, from tests of fundamental physics such as geometric phases and general relativity~\cite{maclaurin_measurable_2012,wood_observation_2020,wood_magnetic_2017,smiciklas_new_2011,terrano_comagnetometer_2021,arditty_sagnac_1981}, to industrial applications such as navigation in the absence of global positioning system~\cite{passaro_gyroscope_2017,escobar-alvarez_r-advance_2018,fang_metrology_2016,zhao_inertial_2022}. In recent years, sensing technologies based on quantum systems have shown promising performances in sensitivity, resolution, stability, etc. compared with their classical counterparts~\cite{degen_quantum_2017}. 
Among these systems, nuclear spins in solid-state platforms,  in particular nitrogen-vacancy (NV) centers in diamonds~\cite{doherty_nitrogen-vacancy_2013}, have become an attractive candidate due to their long coherence times, ambient operation conditions, small size, and fabrication capabilities~\cite{ajoy_stable_2012,jaskula_cross-sensor_2019,jarmola_demonstration_2021,soshenko_nuclear_2021,wood_quantum_2018,wang_characterizing_2023}.

An inertial measurement requires the system to evolve (e.g. rotate) relative to an inertial reference frame. Existing NV gyroscopes are mostly based on the nuclear spin of the $^{14}$N ($I=1$) atom~\cite{ajoy_stable_2012,jaskula_cross-sensor_2019,jarmola_demonstration_2021,soshenko_nuclear_2021} whose nuclear quadrupole term quantizes the nuclear spin along the NV orientation $\hat{z}_{\text{NV}}$ and constrains the nuclear spin to detect longitudinal rotations about $\hat{z}_{\text{NV}}$, that is, the crystal [111] direction. The rotation rate can be extracted from the accumulated dynamic phase using a Ramsey sequence. In contrast, the nuclear spin of the $^{15}$N atom is a spin-$\frac{1}{2}$ without a quadrupole term and thus is released from such a constraint. Although the hyperfine interaction with the NV electronic spin still effectively applies a large field that constrains the quantization along $\hat{z}_{\text{NV}}$, such a field vanishes for the NV spin state $m_S=0$. 
Thus, the nuclear spin can be isolated from the crystal orientation and makes the system an ideal platform for inertial sensing. Moreover, $^{15}$N nuclear spin has only two energy levels, making it easier to polarize, control, and read out. Despite these benefits, the use of $^{15}$N in NV gyroscopes remains less explored and the current sensing protocols, limited to the conventional Ramsey methods, are still limited by the nuclear spin dephasing time  $T_{2n}^*$ ~\cite{burgler_all-optical_2023}, which is degraded by the presence of transverse field inhomogeneities~\cite{wang_emulated_2023}.

In this work, we develop a gyroscope protocol based on the $^{15}$N nuclear spin in diamond NV centers. Instead of using a superposition state to measure longitudinal rotations about $\hat{z}_{\text{NV}}$, we now use the population state to measure rotations along a transverse direction, thus being robust against spin dephasing. 
While the NV spin quantization axis co-rotates with the diamond, the nuclear spin either adiabatically follows its eigenstate under a slow rotation (Fig \ref{fig:1}(b)), or remains in its initial state under a fast rotation (Fig \ref{fig:1}(c)), both yielding relative rotations between the two spins that can be used to extract the rotation rate, covering a broad dynamic range. In particular, the NV-nuclear hyperfine interaction enhances the transverse Zeeman coupling of the nuclear spin by a (magnetic field dependent) factor of 15 to a few thousand, which can be used to boost the nuclear spin rotation rate and thus the gyroscope sensitivity by a same factor. Numerical simulations demonstrate that our gyroscope protocol is robust against magnetic noise and only limited by the NV lifetime $T_{1e}$. Besides rotation sensing, our gyroscope can provide insights into experimental tests of fundamental physics such as Lorentz invariance, relativistic geometric phases, and the Einstein de Haas effect~\cite{stedman1997ring, maclaurin_measurable_2012, jaafar2009dynamics}.

\begin{figure*}[htbp]
\includegraphics[width=\textwidth]{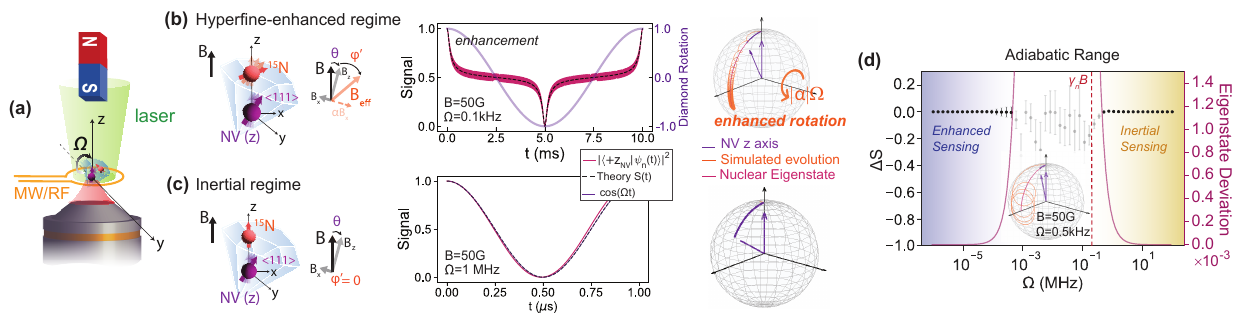}
\caption{\label{fig:1} \textbf{Gyroscope principle.} (a) Gyroscope based on an electro-nuclear spin system composed of NV electronic and nuclear spins in diamond. (b) Hyperfine-enhanced rotation regime ($\Omega \ll \gamma_n B$).  The relative rotation between the NV and the nuclear spin is enhanced due to the hyperfine interaction.  For a diamond rotation rate $\Omega = (2\pi)0.1$kHz and magnetic field $B = 50$G, the measured overlap of the final electronic and nuclear spin states (referenced with the diamond rotation) is simulated, with an accompanying schematic showing the NV $\hat{z}_{\text{NV}}$ axis rotation, nuclear spin evolution, and nuclear spin eigenstate. (c) Inertial regime ($\Omega \gg \gamma_n B$). The nuclear spin remains in its initial state and the relative rotation corresponds directly to the rotation of the diamond.  The simulated signal and system evolution is shown for a fast rotation $\Omega = (2 \pi) 1$ MHz. (d) Signal deviation from the ideal regimes presented in (b) and (c) and the corresponding eigenstate deviation of the nuclear spin Hamiltonian. In the intermediate regime, it is hard to attribute the population signal to either hyperfine-enhanced or inertial regimes.}
\end{figure*}

\textit{System.---}
The physical system of our gyroscope is based on an electro-nuclear spin system hosted by point defects in solid-state platforms. Specifically, we use NV centers in diamond with enriched $^{15}$N isotope to illustrate the protocol. 
The NV center consists of an electronic spin-1 and a nitrogen nuclear spin-$\frac{1}{2}$. Due to its isolation from the environment, the nuclear spin is used as the inertial sensor, while the electronic spin is used for initializing and reading out the nuclear spin state~\cite{doherty_nitrogen-vacancy_2013}. The ground state Hamiltonian of the NV center can be written as
\begin{equation}
\label{eq:Hamiltonian}
    H = \Vec{S}\cdot \bold{D}\cdot \Vec{S}  + \Vec{S} \cdot \bold{A} \cdot \Vec{I} + \gamma_e \Vec{B} \cdot \Vec{S} + \gamma_n \Vec{B} \cdot \Vec{I}
\end{equation}
where $\bold{D}$, $\bold{A}$ are zero-field splitting (ZFS) and hyperfine tensors and $\gamma_e=(2 \pi)2.802$~MHz/G,  $\gamma_n=(2 \pi)0.432$~kHz/G are the gyromagnetic ratios of the electronic and nuclear spin. When the $\hat{z}$ axis is chosen to be along the N-to-V orientation intrinsic to the diamond crystal (henceforth referred to as $\hat z_{NV}$ in the NV frame), both tensors are diagonal: the ZFS tensor has a longitudinal term $D=(2 \pi)2.87$~GHz; the longitudinal and transverse components of the hyperfine tensor are $A_{zz}=(2 \pi)3.03$~MHz and $A_{\perp}=(2 \pi)3.65$~MHz~\cite{felton_hyperfine_2009,lourette_temperature_2022}.

Under an external magnetic field $\Vec{B}$ satisfying $\gamma_eB\ll D$, the electronic spin is quantized along the NV orientation by the large ZFS and is thus tied to the diamond crystal. In contrast, the nuclear spin energy is only defined by the external magnetic field and hyperfine interaction with the electron spin, with an effective Hamiltonian in the NV frame~\cite{chen_measurement_2015,sangtawesin_hyperfine-enhanced_2016,oon_ramsey_2022-1}
\begin{equation}
\label{eq:H_I}
    H_I=\gamma_n\left[B_z{I}_z +\alpha_{m_s}  B_x {I}_x\right]+A_{zz}m_s{I}_z,
\end{equation}
where $m_s$ is the electronic spin Zeeman state and $\alpha_{m_s}$ is an enhancement factor of the transverse Zeeman coupling induced by mixing with the electronic spin states due to the transverse hyperfine interaction~\cite{chen_measurement_2015,sangtawesin_hyperfine-enhanced_2016}. When the external magnetic field is small ($\gamma_e B \ll D$), such a factor can be approximated to a constant $\alpha_{m_s}=(1-2\kappa+3\kappa m_s^2)$ with $\kappa\approx\gamma_eA_\perp/(\gamma_n D)\approx 8.26$~\cite{SOM,oon_ramsey_2022-1}.

Ideally, when there is no external magnetic field and $m_S =  0$, the nuclear spin is effectively decoupled from the NV electronic spin and a physical rotation of the diamond along a transverse direction $\hat{y}$ rotates the electronic spin eigenstates while leaving the nuclear spin unchanged.  When the rotation rate satisfies $\Omega \ll D - \gamma_eB$, the NV electronic spin state adiabatically follows the NV orientation axis $\hat{z}_{\text{NV}}$. In the NV frame, the nuclear spin initialized to $\ket{m_I=+1/2}$ rotates in the $z-x$ plane with a rate $-\Omega$ due to its inertia. The nuclear spin population projected onto the NV axis can be measured by mapping it to the electronic spin state with a CNOT gate (achieved by an electron spin $\pi$-pulse conditionally applied on a nuclear spin state~\cite{SOM}) and then reading out through the fluorescence of the NV center. The readout signal is then $S(t)=(1+\cos(\Omega t))/2$, from which the rotation rate can be extracted.

While this protocol is attractive as it does not require gimbals, working at a zero magnetic field is challenging because of the need for magnetic shielding.
In addition, we expect that working at zero field will make the nuclear spin even more susceptible to magnetic noise, which is usually larger at lower frequencies and can lead to dephasing and depolarization, harmful to the sensitivity. It is thus more practical to develop a general protocol for nonzero magnetic field conditions.

\textit{Hyperfine-enhanced regime.---}
Here we show how adding a magnetic field improves the rotation sensing performance by amplifying the nuclear spin rotation. We assume to apply an external magnetic field $B$ in a static frame, and first consider sensing slow rotations of the diamond, in the regime where $\Omega \ll \gamma_n B$.  
With the diamond initially aligned with $\hat z_{NV}$ along the magnetic field,  both electronic and nuclear spins are initialized in the z-eigenstates, $\ket{m_S=0,m_I=1/2}$. When the diamond starts to rotate along $\hat{y}$ by an angle $\theta=\Omega t$, the ZFS and hyperfine tensors start to rotate. Conversely, in the NV frame, the magnetic field starts to have a transverse component. While the electronic spin is still mainly quantized along $\hat z_{NV}$, the nuclear spin adiabatically follows the effective magnetic field direction $\Vec{B}_{\text{eff}} = -\alpha_0 B\sin\theta \hat{x}_{NV} + B\cos\theta \hat{z}_{NV}$ (Eq.~\eqref{eq:H_I}) (Fig. \ref{fig:1}(b)).   
The effective magnetic field for the nuclear spin rotates with respect to the NV axis by an angle $\varphi^\prime$ satisfying $\tan\varphi^\prime=-\alpha_0 \tan\theta$. For $\theta\ll1$, the nuclear spin rotation in the NV frame is amplified by a factor of $|\varphi^\prime/\theta|\approx |\alpha_0|$, giving an effective rate in the lab frame of $\Omega(1+|\alpha_0|)$. 
The rotation enhancement factor is magnetic-field dependent ~\cite{SOM}, increasing dramatically to its maximum value $\alpha_0 \approx \frac{\gamma_e}{\sqrt{2} \gamma_n} \approx 4.6\times 10^3$ near the GSLAC condition when $\gamma_e B \approx D$ ~\cite{sangtawesin_hyperfine-enhanced_2016,chen_measurement_2015}.  

To extract the rotation rate $\Omega$, we can measure the nuclear spin population along the $\hat{z}_{NV}$ axis by mapping the nuclear spin state onto the NV, yielding a signal
\begin{equation}
\label{eq:S(t)}
  S(t)=\frac{1+\cos\varphi^\prime(t)}{2}=\frac12 (1+\frac{\cos\Omega t}{\sqrt{\cos^2\Omega t+\alpha_0^2\sin^2\Omega t}})
\end{equation} 
In Fig.~\ref{fig:1}(b), we numerically simulate the evolution of the gyroscope system under a rotation rate $\Omega=(2\pi)100$~Hz with an external magnetic field $B =50$~G. The evolution of the NV electronic and nuclear spins is separated out by tracing out the other spin. The simulated nuclear spin population along $z_{NV}$ is shown by the pink curve in Fig.~\ref{fig:1}(b), which matches the theoretical prediction from Eq.~\eqref{eq:S(t)} shown by the black dashed curve.  We note that the fast oscillations around the predicted evolution observed in simulations are a result of the nuclear spin precession about the effective magnetic field. As a reference, the diamond rotation is shown by the purple curve.

In Fig.~\ref{fig:2}(a), we simulate the rotation enhancement factor by explicitly calculating the rotation rate of the nuclear spin. The maximum enhancement is achieved when the magnetic field is aligned with the NV axis, where however the signal in Eq~\eqref{eq:S(t)} doesn't depend on $\alpha_0$. We thus instead 
calculate the rate of change in the nuclear spin population along $\hat{x}$ which is proportional to the rotation angle when it is small (see inset of Fig.~\ref{fig:2}(a)). The simulation results match the theoretically predicted enhancement factor to first order $|\alpha_0|\approx \frac{\gamma_e}{\gamma_n}A_\perp/(D-\gamma_e B)$. 

\begin{figure}[htbp]
\includegraphics[width=0.49\textwidth]{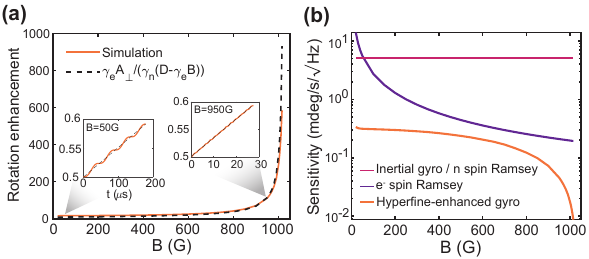}
\caption{\label{fig:2} \textbf{Simulation of the gyroscope protocol.}  (a) Simulated rotation enhancement factor for small $\theta \ll 1$.  The insets show the time evolution of the nuclear spin population along $\hat{x}_{NV}$ used to extract the rotation rate under two exemplary conditions with small and large magnetic fields.  (b) Sensitivity comparison of different gyroscope protocols. For a fair comparison, here we use the same readout efficiency $C=2\%$ across protocols.  For all three nuclear spin-based protocols, we set the signal decay times to 1.5$T_{1e}=7.5$~ms and dead time to $t_d=0.5$~ms. The ``e spin" protocol utilizes the NV electronic spin resonance frequency shifts measured by a Ramsey sequence to sense the rotation rate (see supplement~\cite{SOM} for details), where we set a typical spin dephasing time $T_{2e}^*=0.7$~$\mu$s and $t_d=0.05$~ms~\cite{wang_emulated_2023}. }
\end{figure}

When the nuclear Larmor precession frequency is comparable to the rotation rate, $\Omega \sim \gamma_nB$, the adiabaticity of the nuclear spin evolution begins to break down (as seen on the Bloch sphere in Fig.~\ref{fig:1}(d) for $\Omega = (2\pi)0.5$kHz), and it no longer follows the effective magnetic field.  To further study such effects, we compare simulations over different rotation rates for a fixed magnetic field in Fig.~\ref{fig:1}(d). 
For each rotation rate $\Omega \le \gamma_n B$, we compute the average deviation from the predicted signal in Eq.~\eqref{eq:S(t)} on the left y-axis and compute the non-adiabaticity from the  eigenstate deviation $|\langle +\frac12| \frac{d H_I(t)}{dt}| -\frac12 \rangle|^2$ (where $|m_I = \pm 1/2\rangle$ are the nuclear spin eigenstates) on the right y-axis ~\cite{SOM}.  As the rotation rate increases, the eigenstate deviation increases and after a threshold value, it becomes difficult to directly extract the rotation rate from the measured signal.  To maintain the adiabaticity and sense faster rotations in the hyperfine-enhanced regime, one can increase the bias magnetic field to increase the nuclear spin energy gap.  Thus, quantifying the nuclear eigenstate deviation allows us to define a bespoke maximum rotation rate that one can detect for a given magnetic field, dependent on the desired precision of the signal.

\textit{Inertial regime.---}
While the enhanced rotation factor is limited to sensing a rotation rate that is much less than the nuclear spin energy gap, the proposed gyroscope protocol can still sense rotation rates in the regime where $\Omega \gg \gamma_n B$.  Here, the electron spin is still quantized in the crystal's $\hat{z}_{NV}$ direction, thus following the diamond rotation adiabatically. The effective external magnetic field on the nuclear spin rapidly oscillates and averages to zero similar to the $B=0$ case, so the nuclear spin remains in its initial state,  effectively decoupled from the electron spin. In this scenario, the relative rotation observed between the two spins corresponds directly to the diamond's rotation, and thus in the gyroscope protocol the nuclear spin population signal is given by
\begin{equation}
\label{eq: S(t)_fast}
S(t)=\frac12(1+\cos\theta)=\frac12\left(1+\cos(\Omega t)\right).
\end{equation}
We simulate the evolution of our system under a fast rotation $\Omega = (2\pi)1$ MHz in Fig.~\ref{fig:1}(c) yielding a signal consistent with that predicted by Eq.~\eqref{eq: S(t)_fast}.  Similar to the hyperfine-enhanced regime, calculating the eigenstate deviation (in Fig.~\ref{fig:1}(d)) identifies the minimum rotation rate required for high-fidelity measurements in this regime. 

Thus, combining both regimes, our gyroscope measures a broad dynamic range  $\Omega\lesssim D-\gamma_eB$, except for a small window near the nuclear spin energy gap $\Omega \sim \gamma_n B$. To maximize the enhanced rotation rate in the hyperfine-enhanced regime, one can choose to set the bias magnetic field strength as close as possible to GSLAC.  Nevertheless, at high fields, misalignment of the electron spin from the magnetic field, as well as its small energy gap, can make initialization and readout of the protocol difficult ~\cite{SOM}.  In the following, we discuss and analyze such limits to the sensitivity of the gyroscope, including the system's ultimate lifetime and optimal performance.

\textit{Gyroscope Performance.---}
Quantum sensors' performance is bound by their coherence or relaxation times, which limits the sensing time $t$.   In the enhanced rotation regime, the nuclear spin adiabatically follows its eigenstate and thus its decay induced by an external bath follows a spin relaxation process, which typically yields a relaxation time $T_{1n}$  much longer than the dephasing time $T_{2n}^*$.  
However, the NV electronic spin relaxation $T_{1e}$ has been shown to be a source of decoherence, typically limiting the nuclear spin dephasing, $T_{2n}^*=1.5T_{1e}$~\cite{chen_protecting_2018}). 
Under the gyroscope rotation, an NV spin flip due to relaxation changes the nuclear spin quantization axis (Eq.~\eqref{eq:H_I}), leading to fast decay of what was previously an eigenstate. 
We numerically simulate this process using the Lindblad equation with Lindblad operators  $L_k=\sqrt{\Gamma}\ket{m_s}\bra{m_s^\prime}$, where $m_s, m_s^\prime\in\{-1,0,+1\}$ and the jump rate is $\Gamma=1/(3T_{1e})$. 
Fig.~\ref{fig:4}(a)i shows that in the slow rotation regime the gyroscope signal decay time is $T_{1n}\approx 1.5T_{1e}$. In the inertial regime under fast rotations shown in Fig.~\ref{fig:4}(b)ii, the simulation yields a decay time $T_{1n} > 1.5T_{1e}$. This increased decay time might originate from effective dynamical decoupling effects due to the electron spin's rapid rotation~\cite{SOM,chen_protecting_2018}.  

\begin{figure}[htbp]
\includegraphics[width=0.49\textwidth]{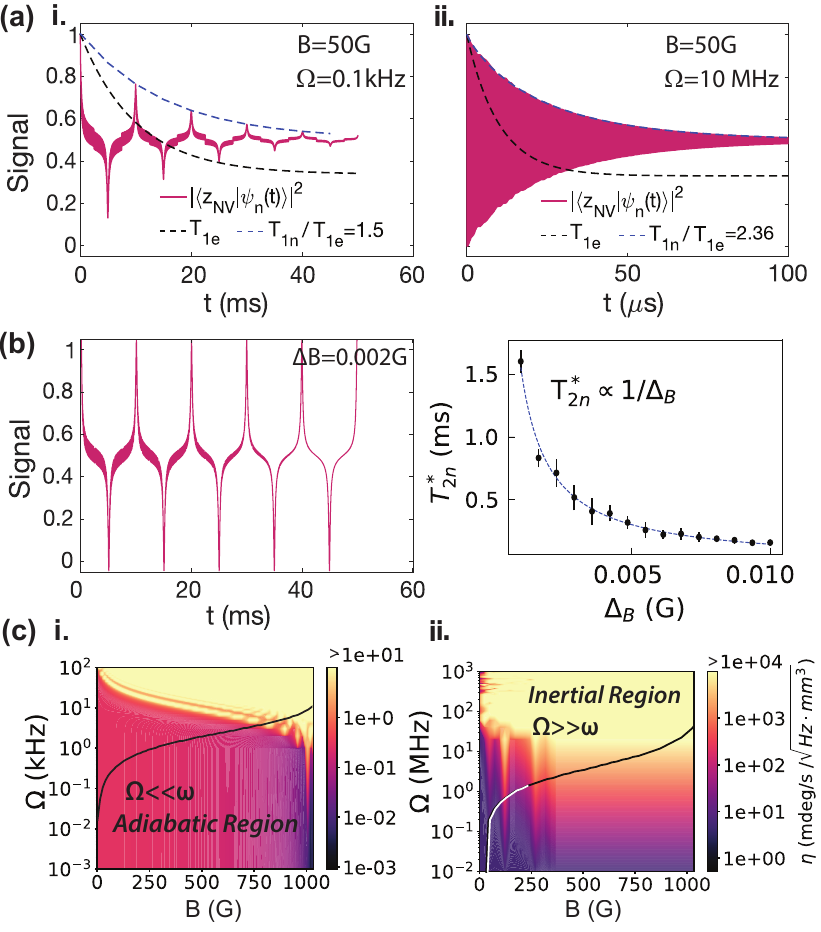}
\caption{\label{fig:4} \textbf{Performance analysis.} (a) Gyroscope signal under electronic spin T1 relaxation at $B = 50$~G in (i) hyperfine-enhanced regime with $\Omega = (2\pi)0.1 $ kHz $\ll \gamma_nB$ and (ii) inertial regime with $\Omega = (2\pi)10 $ MHz.  The decay of the nuclear spin signal envelope is fit to $S(t) = c_0 + c_1e^{-t/\tau}$ for regime (i) and $S(t) = c_0 + c_1e^{-(t/\tau)^{c_2}}\cos(\Omega t)$ for regime (ii) to obtain the decay time $T_{1n}$. (b) Gyroscope signal under static magnetic noise at $B=50$~G and $\Omega=(2\pi)0.1$~kHz.  The variance of the magnetic noise is set to be 0.002~G. The inset shows the dephasing time $T_{2n}^*$ as a function of the noise strength. (c) Sensitivity bound $\eta (\Omega, B)$ for a spin density $n_{NV}/4 \approx 2.5 \times 10^{14}$/mm$^3$. Solid lines indicate the maximum and minimum rotation rates such that the theoretical signal predictions remain valid.}  
\end{figure}

The decay due to the electronic spin relaxation process sets the upper limit of the system coherence, but other noise sources such as magnetic noise induced by the spin bath could dominate, especially in dense spin ensembles. We further explore these effects by simulating the gyroscope signal under magnetic field inhomogeneities in Fig.~\ref{fig:4}(b). In the frame of the nuclear spin, the longitudinal magnetic noise causes a $T_2^*$ dephasing process that results in a decay in its transverse spin components. Intuitively, this might lead to a decay of the system coherence as well; however, even though the fast oscillation decays with time, the main gyroscope signal remains relatively robust against the magnetic noise.   Such an effect may even improve the performance of our gyroscope by suppressing the undesired fast oscillation signal that originates from the spin precession around the effective magnetic field.  

Thus, the gyroscope signal decay is ultimately limited by the  lifetime $T_{1n}$, giving a lower bound of the achievable sensitivity of the gyroscope~\cite{SOM},
\begin{equation}
\label{eq:sensitivity}
    \eta \geq \frac{\sqrt{2}\gamma_n}{\gamma_e}  \frac{e^{t/T_{1n}}}{C \sqrt{N}}\frac{\sqrt{t + t_d}}{t},
\end{equation}
for the hyperfine-enhanced regime, where $t$ is the total sensing time and $t_d$ is the dead time for initialization and readout.  With typical ensemble parameters and conditions (readout efficiency $C \sim 2\%$ \cite{barry_sensitivity_2020}), for a volume $V = 1$mm$^3$ and $N = n_{\text{NV}}V/4 \approx 2.5 \times 10^{14}$ sensor spins and $T_{1n} \sim 7.5$ ms, the sensitivity limit reaches $\eta \sim 1 \times 10^{-3}$(mdeg/s)/$\sqrt{\text{Hz}}$ near GSLAC.  In the inertial regime, the sensitivity follows Eq.~\eqref{eq:sensitivity} but without the enhancement factor $\sim \gamma_n/(\sqrt{2}\gamma_e)$.

Nevertheless, implementation of the gyroscope requires the consideration of many practical limitations;  
here, we briefly discuss these factors and their influence on the sensitivity. 
To take advantage of the hyperfine-enhanced rotation rate, the time for signal readout should be chosen such that the two spins are nearly aligned upon measurement. In particular, it is a compromise between choosing a longer sensing time for optimizing sensitivity and making the measurement as quickly as possible to maintain the enhancement which decreases with the initial rotation angle. Furthermore, under large magnetic fields, optical polarization of the NV spin is difficult with misalignment from the magnetic field. Thus, to operate the gyroscope continuously, both the dead time and sensing time are limited to integer multiples of the rotation period $t + t_d \approx k\pi/\Omega$, which also guarantees that the NV is nearly aligned with the magnetic field during initialization and readout.  For these calibrations, an estimate of the rotation rate needs to be initially acquired (e.g., by using the NV spin transition frequency shift during the rotation). Different from typical Ramsey magnetometry, the sensitivity of our gyroscope depends on the relative angle between two spins in the final state, thus the actual sensitivity needs to be calculated taking these into account (see supplement~\cite{SOM} for details). 

Moreover, during the rotation, initialization and readout of the nuclear spin require selective control of the electronic spin such as $\pi$-pulses only for a specific nuclear spin state.  Thus, for large magnetic fields and fast rotation rates, it is important to consider the resonance frequency shift during the rotation, which affects the control fidelity. 
We conceptually characterize these effects by adding a pulse fidelity factor in the readout efficiency assuming a pulse duration $\sim 0.5 \mu$s in Fig.~\ref{fig:4}(c), where we compute the lower bounded sensitivity of the gyroscope values of $(\Omega, B)$.  Despite these restrictions, these effects can be improved by optimal control techniques or adaptive methods. Thus, our gyroscope remains a versatile platform capable of reaching extremely good sensitivity for bespoke rotation measurements.   

We note that with the assistance of an external magnetic field, the NV electronic spin transition frequency can also be used to extract the rotation rate for a transverse rotation, which can be estimated by Ramsey experiments limited by a dephasing time $T_{2e}^*$. In a spin ensemble, the electron spin dephasing is short $T_{2e}^*\ll T_{1e}$~\cite{wang_emulated_2023}, which leads to a worse sensitivity than the nuclear spin gyroscopes as shown in Fig.~\ref{fig:2}(b).

\textit{Conclusion and Discussion.---}
We propose a novel gyroscope protocol based on $^{15}$NV centers in diamond.  In comparison to conventional Ramsey-type gyroscopes limited to sensing a $\hat{z}$ rotation and suffering from spin dephasing, our protocol uses the more robust population state to sense a transverse rotation.  In particular, when an external magnetic field is fixed in the lab frame, the nuclear spin rotation rate can be significantly enhanced by its hyperfine interaction with the electronic spin, achieving a sensitivity improvement of up to three orders of magnitude. The protocol is robust against magnetic noises and is only limited by $T_1$ relaxation times. 

While our analysis shows that nuclear spin lifetime $T_{1n}$ is still limited by the electron spin $T_{1e}$ relaxation time, which is also the ultimate limit  to the coherence limit of the conventional Ramsey-type gyroscope, our protocol is intrinsically more robust to other sources of dephasing noise that dominate Ramsey gyroscopes and an additional,  significant sensitivity improvement derives from the rotation enhancement. The sensitivity could be further improved by designing dynamical decoupling techniques to extend the nuclear spin lifetime beyond $T_{1e}$ by canceling the deleterious effects of the electronic spin relaxation process~\cite{chen_protecting_2018}.

In addition to building a highly sensitive and compact gyroscope under ambient conditions competitive with atomic gyroscopes, our gyroscope can also exploit the electro-nuclear spin system to provide insights into testing fundamental physics.  
Thus, our gyroscope proves to be a robust and versatile device, with broad opportunities for integration and applications. 

\acknowledgements
This work was supported in part by DARPA DRINQS program (Cooperative Agreement No. D18AC00024). G.W. thanks MathWorks for their support in the form of a Graduate Student Fellowship. The opinions and views expressed in this publication are from the authors and
not necessarily from MathWorks.


\bibliography{main_text}

\begin{thebibliography}{32}%
\makeatletter
\providecommand \@ifxundefined [1]{%
 \@ifx{#1\undefined}
}%
\providecommand \@ifnum [1]{%
 \ifnum #1\expandafter \@firstoftwo
 \else \expandafter \@secondoftwo
 \fi
}%
\providecommand \@ifx [1]{%
 \ifx #1\expandafter \@firstoftwo
 \else \expandafter \@secondoftwo
 \fi
}%
\providecommand \natexlab [1]{#1}%
\providecommand \enquote  [1]{``#1''}%
\providecommand \bibnamefont  [1]{#1}%
\providecommand \bibfnamefont [1]{#1}%
\providecommand \citenamefont [1]{#1}%
\providecommand \href@noop [0]{\@secondoftwo}%
\providecommand \href [0]{\begingroup \@sanitize@url \@href}%
\providecommand \@href[1]{\@@startlink{#1}\@@href}%
\providecommand \@@href[1]{\endgroup#1\@@endlink}%
\providecommand \@sanitize@url [0]{\catcode `\\12\catcode `\$12\catcode
  `\&12\catcode `\#12\catcode `\^12\catcode `\_12\catcode `\%12\relax}%
\providecommand \@@startlink[1]{}%
\providecommand \@@endlink[0]{}%
\providecommand \url  [0]{\begingroup\@sanitize@url \@url }%
\providecommand \@url [1]{\endgroup\@href {#1}{\urlprefix }}%
\providecommand \urlprefix  [0]{URL }%
\providecommand \Eprint [0]{\href }%
\providecommand \doibase [0]{https://doi.org/}%
\providecommand \selectlanguage [0]{\@gobble}%
\providecommand \bibinfo  [0]{\@secondoftwo}%
\providecommand \bibfield  [0]{\@secondoftwo}%
\providecommand \translation [1]{[#1]}%
\providecommand \BibitemOpen [0]{}%
\providecommand \bibitemStop [0]{}%
\providecommand \bibitemNoStop [0]{.\EOS\space}%
\providecommand \EOS [0]{\spacefactor3000\relax}%
\providecommand \BibitemShut  [1]{\csname bibitem#1\endcsname}%
\let\auto@bib@innerbib\@empty
\bibitem [{\citenamefont {Maclaurin}\ \emph {et~al.}(2012)\citenamefont
  {Maclaurin}, \citenamefont {Doherty}, \citenamefont {Hollenberg},\ and\
  \citenamefont {Martin}}]{maclaurin_measurable_2012}%
  \BibitemOpen
  \bibfield  {author} {\bibinfo {author} {\bibfnamefont {D.}~\bibnamefont
  {Maclaurin}}, \bibinfo {author} {\bibfnamefont {M.~W.}\ \bibnamefont
  {Doherty}}, \bibinfo {author} {\bibfnamefont {L.~C.~L.}\ \bibnamefont
  {Hollenberg}},\ and\ \bibinfo {author} {\bibfnamefont {A.~M.}\ \bibnamefont
  {Martin}},\ }\bibfield  {title} {\bibinfo {title} {Measurable {{Quantum
  Geometric Phase}} from a {{Rotating Single Spin}}},\ }\href
  {https://doi.org/10.1103/PhysRevLett.108.240403} {\bibfield  {journal}
  {\bibinfo  {journal} {Phys. Rev. Lett.}\ }\textbf {\bibinfo {volume} {108}},\
  \bibinfo {pages} {240403} (\bibinfo {year} {2012})}\BibitemShut {NoStop}%
\bibitem [{\citenamefont {Wood}\ \emph {et~al.}(2020)\citenamefont {Wood},
  \citenamefont {Hollenberg}, \citenamefont {Scholten},\ and\ \citenamefont
  {Martin}}]{wood_observation_2020}%
  \BibitemOpen
  \bibfield  {author} {\bibinfo {author} {\bibfnamefont {A.~A.}\ \bibnamefont
  {Wood}}, \bibinfo {author} {\bibfnamefont {L.~C.~L.}\ \bibnamefont
  {Hollenberg}}, \bibinfo {author} {\bibfnamefont {R.~E.}\ \bibnamefont
  {Scholten}},\ and\ \bibinfo {author} {\bibfnamefont {A.~M.}\ \bibnamefont
  {Martin}},\ }\bibfield  {title} {\bibinfo {title} {Observation of a {{Quantum
  Phase}} from {{Classical Rotation}} of a {{Single Spin}}},\ }\href
  {https://doi.org/10.1103/PhysRevLett.124.020401} {\bibfield  {journal}
  {\bibinfo  {journal} {Phys. Rev. Lett.}\ }\textbf {\bibinfo {volume} {124}},\
  \bibinfo {pages} {020401} (\bibinfo {year} {2020})}\BibitemShut {NoStop}%
\bibitem [{\citenamefont {Wood}\ \emph {et~al.}(2017)\citenamefont {Wood},
  \citenamefont {Lilette}, \citenamefont {Fein}, \citenamefont {Perunicic},
  \citenamefont {Hollenberg}, \citenamefont {Scholten},\ and\ \citenamefont
  {Martin}}]{wood_magnetic_2017}%
  \BibitemOpen
  \bibfield  {author} {\bibinfo {author} {\bibfnamefont {A.~A.}\ \bibnamefont
  {Wood}}, \bibinfo {author} {\bibfnamefont {E.}~\bibnamefont {Lilette}},
  \bibinfo {author} {\bibfnamefont {Y.~Y.}\ \bibnamefont {Fein}}, \bibinfo
  {author} {\bibfnamefont {V.~S.}\ \bibnamefont {Perunicic}}, \bibinfo {author}
  {\bibfnamefont {L.~C.~L.}\ \bibnamefont {Hollenberg}}, \bibinfo {author}
  {\bibfnamefont {R.~E.}\ \bibnamefont {Scholten}},\ and\ \bibinfo {author}
  {\bibfnamefont {A.~M.}\ \bibnamefont {Martin}},\ }\bibfield  {title}
  {\bibinfo {title} {Magnetic pseudo-fields in a rotating electron\textendash
  nuclear spin system},\ }\href {https://doi.org/10.1038/nphys4221} {\bibfield
  {journal} {\bibinfo  {journal} {Nat. Phys.}\ }\textbf {\bibinfo {volume}
  {13}},\ \bibinfo {pages} {1070} (\bibinfo {year} {2017})}\BibitemShut
  {NoStop}%
\bibitem [{\citenamefont {Smiciklas}\ \emph {et~al.}(2011)\citenamefont
  {Smiciklas}, \citenamefont {Brown}, \citenamefont {Cheuk}, \citenamefont
  {Smullin},\ and\ \citenamefont {Romalis}}]{smiciklas_new_2011}%
  \BibitemOpen
  \bibfield  {author} {\bibinfo {author} {\bibfnamefont {M.}~\bibnamefont
  {Smiciklas}}, \bibinfo {author} {\bibfnamefont {J.~M.}\ \bibnamefont
  {Brown}}, \bibinfo {author} {\bibfnamefont {L.~W.}\ \bibnamefont {Cheuk}},
  \bibinfo {author} {\bibfnamefont {S.~J.}\ \bibnamefont {Smullin}},\ and\
  \bibinfo {author} {\bibfnamefont {M.~V.}\ \bibnamefont {Romalis}},\
  }\bibfield  {title} {\bibinfo {title} {New {{Test}} of {{Local Lorentz
  Invariance Using}} a {{Ne}} 21 - {{Rb}} - {{K Comagnetometer}}},\ }\href
  {https://doi.org/10.1103/PhysRevLett.107.171604} {\bibfield  {journal}
  {\bibinfo  {journal} {Phys. Rev. Lett.}\ }\textbf {\bibinfo {volume} {107}},\
  \bibinfo {pages} {171604} (\bibinfo {year} {2011})}\BibitemShut {NoStop}%
\bibitem [{\citenamefont {Terrano}\ and\ \citenamefont
  {Romalis}(2021)}]{terrano_comagnetometer_2021}%
  \BibitemOpen
  \bibfield  {author} {\bibinfo {author} {\bibfnamefont {W.~A.}\ \bibnamefont
  {Terrano}}\ and\ \bibinfo {author} {\bibfnamefont {M.~V.}\ \bibnamefont
  {Romalis}},\ }\bibfield  {title} {\bibinfo {title} {Comagnetometer probes of
  dark matter and new physics},\ }\href
  {https://doi.org/10.1088/2058-9565/ac1ae0} {\bibfield  {journal} {\bibinfo
  {journal} {Quantum Sci. Technol.}\ }\textbf {\bibinfo {volume} {7}},\
  \bibinfo {pages} {014001} (\bibinfo {year} {2021})}\BibitemShut {NoStop}%
\bibitem [{\citenamefont {Arditty}\ and\ \citenamefont
  {Lef{\`e}vre}(1981)}]{arditty_sagnac_1981}%
  \BibitemOpen
  \bibfield  {author} {\bibinfo {author} {\bibfnamefont {H.~J.}\ \bibnamefont
  {Arditty}}\ and\ \bibinfo {author} {\bibfnamefont {H.~C.}\ \bibnamefont
  {Lef{\`e}vre}},\ }\bibfield  {title} {\bibinfo {title} {Sagnac effect in
  fiber gyroscopes},\ }\href {https://doi.org/10.1364/OL.6.000401} {\bibfield
  {journal} {\bibinfo  {journal} {Opt. Lett.}\ }\textbf {\bibinfo {volume}
  {6}},\ \bibinfo {pages} {401} (\bibinfo {year} {1981})}\BibitemShut {NoStop}%
\bibitem [{\citenamefont {Passaro}\ \emph {et~al.}(2017)\citenamefont
  {Passaro}, \citenamefont {Cuccovillo}, \citenamefont {Vaiani}, \citenamefont
  {De~Carlo},\ and\ \citenamefont {Campanella}}]{passaro_gyroscope_2017}%
  \BibitemOpen
  \bibfield  {author} {\bibinfo {author} {\bibfnamefont {V.~M.~N.}\
  \bibnamefont {Passaro}}, \bibinfo {author} {\bibfnamefont {A.}~\bibnamefont
  {Cuccovillo}}, \bibinfo {author} {\bibfnamefont {L.}~\bibnamefont {Vaiani}},
  \bibinfo {author} {\bibfnamefont {M.}~\bibnamefont {De~Carlo}},\ and\
  \bibinfo {author} {\bibfnamefont {C.~E.}\ \bibnamefont {Campanella}},\
  }\bibfield  {title} {\bibinfo {title} {Gyroscope {{Technology}} and
  {{Applications}}: {{A Review}} in the {{Industrial Perspective}}},\ }\href
  {https://doi.org/10.3390/s17102284} {\bibfield  {journal} {\bibinfo
  {journal} {Sensors}\ }\textbf {\bibinfo {volume} {17}},\ \bibinfo {pages}
  {2284} (\bibinfo {year} {2017})}\BibitemShut {NoStop}%
\bibitem [{\citenamefont {{Escobar-Alvarez}}\ \emph {et~al.}(2018)\citenamefont
  {{Escobar-Alvarez}}, \citenamefont {Johnson}, \citenamefont {Hebble},
  \citenamefont {Klingebiel}, \citenamefont {Quintero}, \citenamefont
  {Regenstein},\ and\ \citenamefont
  {Browning}}]{escobar-alvarez_r-advance_2018}%
  \BibitemOpen
  \bibfield  {author} {\bibinfo {author} {\bibfnamefont {H.~D.}\ \bibnamefont
  {{Escobar-Alvarez}}}, \bibinfo {author} {\bibfnamefont {N.}~\bibnamefont
  {Johnson}}, \bibinfo {author} {\bibfnamefont {T.}~\bibnamefont {Hebble}},
  \bibinfo {author} {\bibfnamefont {K.}~\bibnamefont {Klingebiel}}, \bibinfo
  {author} {\bibfnamefont {S.~A.~P.}\ \bibnamefont {Quintero}}, \bibinfo
  {author} {\bibfnamefont {J.}~\bibnamefont {Regenstein}},\ and\ \bibinfo
  {author} {\bibfnamefont {N.~A.}\ \bibnamefont {Browning}},\ }\bibfield
  {title} {\bibinfo {title} {R-{{ADVANCE}}: {{Rapid Adaptive Prediction}} for
  {{Vision-based Autonomous Navigation}}, {{Control}}, and {{Evasion}}},\
  }\href {https://doi.org/10.1002/rob.21744} {\bibfield  {journal} {\bibinfo
  {journal} {J. Field Robot.}\ }\textbf {\bibinfo {volume} {35}},\ \bibinfo
  {pages} {91} (\bibinfo {year} {2018})}\BibitemShut {NoStop}%
\bibitem [{\citenamefont {Fang}\ \emph {et~al.}(2016)\citenamefont {Fang},
  \citenamefont {Dutta}, \citenamefont {Gillot}, \citenamefont {Savoie},
  \citenamefont {Lautier}, \citenamefont {Cheng}, \citenamefont {Alzar},
  \citenamefont {Geiger}, \citenamefont {Merlet}, \citenamefont {Santos},\ and\
  \citenamefont {Landragin}}]{fang_metrology_2016}%
  \BibitemOpen
  \bibfield  {author} {\bibinfo {author} {\bibfnamefont {B.}~\bibnamefont
  {Fang}}, \bibinfo {author} {\bibfnamefont {I.}~\bibnamefont {Dutta}},
  \bibinfo {author} {\bibfnamefont {P.}~\bibnamefont {Gillot}}, \bibinfo
  {author} {\bibfnamefont {D.}~\bibnamefont {Savoie}}, \bibinfo {author}
  {\bibfnamefont {J.}~\bibnamefont {Lautier}}, \bibinfo {author} {\bibfnamefont
  {B.}~\bibnamefont {Cheng}}, \bibinfo {author} {\bibfnamefont {C.~L.~G.}\
  \bibnamefont {Alzar}}, \bibinfo {author} {\bibfnamefont {R.}~\bibnamefont
  {Geiger}}, \bibinfo {author} {\bibfnamefont {S.}~\bibnamefont {Merlet}},
  \bibinfo {author} {\bibfnamefont {F.~P.~D.}\ \bibnamefont {Santos}},\ and\
  \bibinfo {author} {\bibfnamefont {A.}~\bibnamefont {Landragin}},\ }\bibfield
  {title} {\bibinfo {title} {Metrology with {{Atom Interferometry}}: {{Inertial
  Sensors}} from {{Laboratory}} to {{Field Applications}}},\ }\href
  {https://doi.org/10.1088/1742-6596/723/1/012049} {\bibfield  {journal}
  {\bibinfo  {journal} {J. Phys. Conf. Ser.}\ }\textbf {\bibinfo {volume}
  {723}},\ \bibinfo {pages} {012049} (\bibinfo {year} {2016})}\BibitemShut
  {NoStop}%
\bibitem [{\citenamefont {Zhao}\ \emph {et~al.}(2022)\citenamefont {Zhao},
  \citenamefont {Shen}, \citenamefont {Ji},\ and\ \citenamefont
  {Huang}}]{zhao_inertial_2022}%
  \BibitemOpen
  \bibfield  {author} {\bibinfo {author} {\bibfnamefont {L.}~\bibnamefont
  {Zhao}}, \bibinfo {author} {\bibfnamefont {X.}~\bibnamefont {Shen}}, \bibinfo
  {author} {\bibfnamefont {L.}~\bibnamefont {Ji}},\ and\ \bibinfo {author}
  {\bibfnamefont {P.}~\bibnamefont {Huang}},\ }\bibfield  {title} {\bibinfo
  {title} {Inertial measurement with solid-state spins of nitrogen-vacancy
  center in diamond},\ }\href {https://doi.org/10.1080/23746149.2021.2004921}
  {\bibfield  {journal} {\bibinfo  {journal} {Adv. Phys.: X}\ }\textbf
  {\bibinfo {volume} {7}},\ \bibinfo {pages} {2004921} (\bibinfo {year}
  {2022})}\BibitemShut {NoStop}%
\bibitem [{\citenamefont {Degen}\ \emph {et~al.}(2017)\citenamefont {Degen},
  \citenamefont {Reinhard},\ and\ \citenamefont
  {Cappellaro}}]{degen_quantum_2017}%
  \BibitemOpen
  \bibfield  {author} {\bibinfo {author} {\bibfnamefont {C.~L.}\ \bibnamefont
  {Degen}}, \bibinfo {author} {\bibfnamefont {F.}~\bibnamefont {Reinhard}},\
  and\ \bibinfo {author} {\bibfnamefont {P.}~\bibnamefont {Cappellaro}},\
  }\bibfield  {title} {\bibinfo {title} {Quantum sensing},\ }\href
  {https://doi.org/10.1103/RevModPhys.89.035002} {\bibfield  {journal}
  {\bibinfo  {journal} {Rev. Mod. Phys.}\ }\textbf {\bibinfo {volume} {89}},\
  \bibinfo {pages} {035002} (\bibinfo {year} {2017})}\BibitemShut {NoStop}%
\bibitem [{\citenamefont {Doherty}\ \emph {et~al.}(2013)\citenamefont
  {Doherty}, \citenamefont {Manson}, \citenamefont {Delaney}, \citenamefont
  {Jelezko}, \citenamefont {Wrachtrup},\ and\ \citenamefont
  {Hollenberg}}]{doherty_nitrogen-vacancy_2013}%
  \BibitemOpen
  \bibfield  {author} {\bibinfo {author} {\bibfnamefont {M.~W.}\ \bibnamefont
  {Doherty}}, \bibinfo {author} {\bibfnamefont {N.~B.}\ \bibnamefont {Manson}},
  \bibinfo {author} {\bibfnamefont {P.}~\bibnamefont {Delaney}}, \bibinfo
  {author} {\bibfnamefont {F.}~\bibnamefont {Jelezko}}, \bibinfo {author}
  {\bibfnamefont {J.}~\bibnamefont {Wrachtrup}},\ and\ \bibinfo {author}
  {\bibfnamefont {L.~C.}\ \bibnamefont {Hollenberg}},\ }\bibfield  {title}
  {\bibinfo {title} {The nitrogen-vacancy colour centre in diamond},\ }\href
  {https://doi.org/10.1016/j.physrep.2013.02.001} {\bibfield  {journal}
  {\bibinfo  {journal} {Phys. Rep.}\ }\textbf {\bibinfo {volume} {528}},\
  \bibinfo {pages} {1} (\bibinfo {year} {2013})}\BibitemShut {NoStop}%
\bibitem [{\citenamefont {Ajoy}\ and\ \citenamefont
  {Cappellaro}(2012)}]{ajoy_stable_2012}%
  \BibitemOpen
  \bibfield  {author} {\bibinfo {author} {\bibfnamefont {A.}~\bibnamefont
  {Ajoy}}\ and\ \bibinfo {author} {\bibfnamefont {P.}~\bibnamefont
  {Cappellaro}},\ }\bibfield  {title} {\bibinfo {title} {Stable three-axis
  nuclear-spin gyroscope in diamond},\ }\href
  {https://doi.org/10.1103/PhysRevA.86.062104} {\bibfield  {journal} {\bibinfo
  {journal} {Phy. Rev. A}\ }\textbf {\bibinfo {volume} {86}},\ \bibinfo {pages}
  {062104} (\bibinfo {year} {2012})}\BibitemShut {NoStop}%
\bibitem [{\citenamefont {Jaskula}\ \emph {et~al.}(2019)\citenamefont
  {Jaskula}, \citenamefont {Saha}, \citenamefont {Ajoy}, \citenamefont
  {Twitchen}, \citenamefont {Markham},\ and\ \citenamefont
  {Cappellaro}}]{jaskula_cross-sensor_2019}%
  \BibitemOpen
  \bibfield  {author} {\bibinfo {author} {\bibfnamefont {J.-C.}\ \bibnamefont
  {Jaskula}}, \bibinfo {author} {\bibfnamefont {K.}~\bibnamefont {Saha}},
  \bibinfo {author} {\bibfnamefont {A.}~\bibnamefont {Ajoy}}, \bibinfo {author}
  {\bibfnamefont {D.}~\bibnamefont {Twitchen}}, \bibinfo {author}
  {\bibfnamefont {M.}~\bibnamefont {Markham}},\ and\ \bibinfo {author}
  {\bibfnamefont {P.}~\bibnamefont {Cappellaro}},\ }\bibfield  {title}
  {\bibinfo {title} {Cross-{{Sensor Feedback Stabilization}} of an {{Emulated
  Quantum Spin Gyroscope}}},\ }\href
  {https://doi.org/10.1103/PhysRevApplied.11.054010} {\bibfield  {journal}
  {\bibinfo  {journal} {Phy. Rev. Applied}\ }\textbf {\bibinfo {volume} {11}},\
  \bibinfo {pages} {054010} (\bibinfo {year} {2019})}\BibitemShut {NoStop}%
\bibitem [{\citenamefont {Jarmola}\ \emph {et~al.}(2021)\citenamefont
  {Jarmola}, \citenamefont {Lourette}, \citenamefont {Acosta}, \citenamefont
  {Birdwell}, \citenamefont {Bl{\"u}mler}, \citenamefont {Budker},
  \citenamefont {Ivanov},\ and\ \citenamefont
  {Malinovsky}}]{jarmola_demonstration_2021}%
  \BibitemOpen
  \bibfield  {author} {\bibinfo {author} {\bibfnamefont {A.}~\bibnamefont
  {Jarmola}}, \bibinfo {author} {\bibfnamefont {S.}~\bibnamefont {Lourette}},
  \bibinfo {author} {\bibfnamefont {V.~M.}\ \bibnamefont {Acosta}}, \bibinfo
  {author} {\bibfnamefont {A.~G.}\ \bibnamefont {Birdwell}}, \bibinfo {author}
  {\bibfnamefont {P.}~\bibnamefont {Bl{\"u}mler}}, \bibinfo {author}
  {\bibfnamefont {D.}~\bibnamefont {Budker}}, \bibinfo {author} {\bibfnamefont
  {T.}~\bibnamefont {Ivanov}},\ and\ \bibinfo {author} {\bibfnamefont {V.~S.}\
  \bibnamefont {Malinovsky}},\ }\bibfield  {title} {\bibinfo {title}
  {Demonstration of diamond nuclear spin gyroscope},\ }\href
  {https://doi.org/10.1126/sciadv.abl3840} {\bibfield  {journal} {\bibinfo
  {journal} {Sci. Adv.}\ }\textbf {\bibinfo {volume} {7}},\ \bibinfo {pages}
  {eabl3840} (\bibinfo {year} {2021})}\BibitemShut {NoStop}%
\bibitem [{\citenamefont {Soshenko}\ \emph {et~al.}(2021)\citenamefont
  {Soshenko}, \citenamefont {Bolshedvorskii}, \citenamefont {Rubinas},
  \citenamefont {Sorokin}, \citenamefont {Smolyaninov}, \citenamefont
  {Vorobyov},\ and\ \citenamefont {Akimov}}]{soshenko_nuclear_2021}%
  \BibitemOpen
  \bibfield  {author} {\bibinfo {author} {\bibfnamefont {V.~V.}\ \bibnamefont
  {Soshenko}}, \bibinfo {author} {\bibfnamefont {S.~V.}\ \bibnamefont
  {Bolshedvorskii}}, \bibinfo {author} {\bibfnamefont {O.}~\bibnamefont
  {Rubinas}}, \bibinfo {author} {\bibfnamefont {V.~N.}\ \bibnamefont
  {Sorokin}}, \bibinfo {author} {\bibfnamefont {A.~N.}\ \bibnamefont
  {Smolyaninov}}, \bibinfo {author} {\bibfnamefont {V.~V.}\ \bibnamefont
  {Vorobyov}},\ and\ \bibinfo {author} {\bibfnamefont {A.~V.}\ \bibnamefont
  {Akimov}},\ }\bibfield  {title} {\bibinfo {title} {Nuclear {{Spin Gyroscope}}
  based on the {{Nitrogen Vacancy Center}} in {{Diamond}}},\ }\href
  {https://doi.org/10.1103/PhysRevLett.126.197702} {\bibfield  {journal}
  {\bibinfo  {journal} {Phys. Rev. Lett.}\ }\textbf {\bibinfo {volume} {126}},\
  \bibinfo {pages} {6} (\bibinfo {year} {2021})}\BibitemShut {NoStop}%
\bibitem [{\citenamefont {Wood}\ \emph {et~al.}(2018)\citenamefont {Wood},
  \citenamefont {Lilette}, \citenamefont {Fein}, \citenamefont {Tomek},
  \citenamefont {McGuinness}, \citenamefont {Hollenberg}, \citenamefont
  {Scholten},\ and\ \citenamefont {Martin}}]{wood_quantum_2018}%
  \BibitemOpen
  \bibfield  {author} {\bibinfo {author} {\bibfnamefont {A.~A.}\ \bibnamefont
  {Wood}}, \bibinfo {author} {\bibfnamefont {E.}~\bibnamefont {Lilette}},
  \bibinfo {author} {\bibfnamefont {Y.~Y.}\ \bibnamefont {Fein}}, \bibinfo
  {author} {\bibfnamefont {N.}~\bibnamefont {Tomek}}, \bibinfo {author}
  {\bibfnamefont {L.~P.}\ \bibnamefont {McGuinness}}, \bibinfo {author}
  {\bibfnamefont {L.~C.~L.}\ \bibnamefont {Hollenberg}}, \bibinfo {author}
  {\bibfnamefont {R.~E.}\ \bibnamefont {Scholten}},\ and\ \bibinfo {author}
  {\bibfnamefont {A.~M.}\ \bibnamefont {Martin}},\ }\bibfield  {title}
  {\bibinfo {title} {Quantum measurement of a rapidly rotating spin qubit in
  diamond},\ }\href {https://doi.org/10.1126/sciadv.aar7691} {\bibfield
  {journal} {\bibinfo  {journal} {Sci. Adv.}\ }\textbf {\bibinfo {volume}
  {4}},\ \bibinfo {pages} {eaar7691} (\bibinfo {year} {2018})}\BibitemShut
  {NoStop}%
\bibitem [{\citenamefont {Wang}\ \emph {et~al.}(2023)\citenamefont {Wang},
  \citenamefont {Barr}, \citenamefont {Tang}, \citenamefont {Chen},
  \citenamefont {Li}, \citenamefont {Xu}, \citenamefont {Stasiuk},
  \citenamefont {Li},\ and\ \citenamefont
  {Cappellaro}}]{wang_characterizing_2023}%
  \BibitemOpen
  \bibfield  {author} {\bibinfo {author} {\bibfnamefont {G.}~\bibnamefont
  {Wang}}, \bibinfo {author} {\bibfnamefont {A.~R.}\ \bibnamefont {Barr}},
  \bibinfo {author} {\bibfnamefont {H.}~\bibnamefont {Tang}}, \bibinfo {author}
  {\bibfnamefont {M.}~\bibnamefont {Chen}}, \bibinfo {author} {\bibfnamefont
  {C.}~\bibnamefont {Li}}, \bibinfo {author} {\bibfnamefont {H.}~\bibnamefont
  {Xu}}, \bibinfo {author} {\bibfnamefont {A.}~\bibnamefont {Stasiuk}},
  \bibinfo {author} {\bibfnamefont {J.}~\bibnamefont {Li}},\ and\ \bibinfo
  {author} {\bibfnamefont {P.}~\bibnamefont {Cappellaro}},\ }\bibfield  {title}
  {\bibinfo {title} {Characterizing temperature and strain variations with
  qubit ensembles for their robust coherence protection},\ }\href
  {https://doi.org/10.1103/PhysRevLett.131.043602} {\bibfield  {journal}
  {\bibinfo  {journal} {Phys. Rev. Lett.}\ }\textbf {\bibinfo {volume} {131}},\
  \bibinfo {pages} {043602} (\bibinfo {year} {2023})}\BibitemShut {NoStop}%
\bibitem [{\citenamefont {B{\"u}rgler}\ \emph {et~al.}(2023)\citenamefont
  {B{\"u}rgler}, \citenamefont {Sjolander}, \citenamefont {Brinza},
  \citenamefont {Tallaire}, \citenamefont {Achard},\ and\ \citenamefont
  {Maletinsky}}]{burgler_all-optical_2023}%
  \BibitemOpen
  \bibfield  {author} {\bibinfo {author} {\bibfnamefont {B.}~\bibnamefont
  {B{\"u}rgler}}, \bibinfo {author} {\bibfnamefont {T.~F.}\ \bibnamefont
  {Sjolander}}, \bibinfo {author} {\bibfnamefont {O.}~\bibnamefont {Brinza}},
  \bibinfo {author} {\bibfnamefont {A.}~\bibnamefont {Tallaire}}, \bibinfo
  {author} {\bibfnamefont {J.}~\bibnamefont {Achard}},\ and\ \bibinfo {author}
  {\bibfnamefont {P.}~\bibnamefont {Maletinsky}},\ }\bibfield  {title}
  {\bibinfo {title} {All-optical nuclear quantum sensing using nitrogen-vacancy
  centers in diamond},\ }\href {https://doi.org/10.1038/s41534-023-00724-6}
  {\bibfield  {journal} {\bibinfo  {journal} {npj Quantum Inf.}\ }\textbf
  {\bibinfo {volume} {9}},\ \bibinfo {pages} {1} (\bibinfo {year}
  {2023})}\BibitemShut {NoStop}%
\bibitem [{\citenamefont {Wang}\ \emph {et~al.}(2024)\citenamefont {Wang},
  \citenamefont {Nguyen}, \citenamefont {de~Quilettes}, \citenamefont {Price},
  \citenamefont {Hu}, \citenamefont {Braje},\ and\ \citenamefont
  {Cappellaro}}]{wang_emulated_2023}%
  \BibitemOpen
  \bibfield  {author} {\bibinfo {author} {\bibfnamefont {G.}~\bibnamefont
  {Wang}}, \bibinfo {author} {\bibfnamefont {M.-T.}\ \bibnamefont {Nguyen}},
  \bibinfo {author} {\bibfnamefont {D.~W.}\ \bibnamefont {de~Quilettes}},
  \bibinfo {author} {\bibfnamefont {E.}~\bibnamefont {Price}}, \bibinfo
  {author} {\bibfnamefont {Z.}~\bibnamefont {Hu}}, \bibinfo {author}
  {\bibfnamefont {D.~A.}\ \bibnamefont {Braje}},\ and\ \bibinfo {author}
  {\bibfnamefont {P.}~\bibnamefont {Cappellaro}},\ }\href@noop {} {\bibinfo
  {title} {Emulated nuclear spin gyroscope with $^{15}$nv centers in diamond}}
  (\bibinfo {year} {2024}),\ \Eprint {https://arxiv.org/abs/2401.01333}
  {arXiv:2401.01333 [quant-ph]} \BibitemShut {NoStop}%
\bibitem [{\citenamefont {Stedman}(1997)}]{stedman1997ring}%
  \BibitemOpen
  \bibfield  {author} {\bibinfo {author} {\bibfnamefont {G.}~\bibnamefont
  {Stedman}},\ }\bibfield  {title} {\bibinfo {title} {Ring-laser tests of
  fundamental physics and geophysics},\ }\href@noop {} {\bibfield  {journal}
  {\bibinfo  {journal} {Reports on progress in physics}\ }\textbf {\bibinfo
  {volume} {60}},\ \bibinfo {pages} {615} (\bibinfo {year} {1997})}\BibitemShut
  {NoStop}%
\bibitem [{\citenamefont {Jaafar}\ \emph {et~al.}(2009)\citenamefont {Jaafar},
  \citenamefont {Chudnovsky},\ and\ \citenamefont
  {Garanin}}]{jaafar2009dynamics}%
  \BibitemOpen
  \bibfield  {author} {\bibinfo {author} {\bibfnamefont {R.}~\bibnamefont
  {Jaafar}}, \bibinfo {author} {\bibfnamefont {E.}~\bibnamefont {Chudnovsky}},\
  and\ \bibinfo {author} {\bibfnamefont {D.}~\bibnamefont {Garanin}},\
  }\bibfield  {title} {\bibinfo {title} {Dynamics of the einstein--de haas
  effect: application to a magnetic cantilever},\ }\href@noop {} {\bibfield
  {journal} {\bibinfo  {journal} {Physical Review B}\ }\textbf {\bibinfo
  {volume} {79}},\ \bibinfo {pages} {104410} (\bibinfo {year}
  {2009})}\BibitemShut {NoStop}%
\bibitem [{\citenamefont {Felton}\ \emph {et~al.}(2009)\citenamefont {Felton},
  \citenamefont {Edmonds}, \citenamefont {Newton}, \citenamefont {Martineau},
  \citenamefont {Fisher}, \citenamefont {Twitchen},\ and\ \citenamefont
  {Baker}}]{felton_hyperfine_2009}%
  \BibitemOpen
  \bibfield  {author} {\bibinfo {author} {\bibfnamefont {S.}~\bibnamefont
  {Felton}}, \bibinfo {author} {\bibfnamefont {A.~M.}\ \bibnamefont {Edmonds}},
  \bibinfo {author} {\bibfnamefont {M.~E.}\ \bibnamefont {Newton}}, \bibinfo
  {author} {\bibfnamefont {P.~M.}\ \bibnamefont {Martineau}}, \bibinfo {author}
  {\bibfnamefont {D.}~\bibnamefont {Fisher}}, \bibinfo {author} {\bibfnamefont
  {D.~J.}\ \bibnamefont {Twitchen}},\ and\ \bibinfo {author} {\bibfnamefont
  {J.~M.}\ \bibnamefont {Baker}},\ }\bibfield  {title} {\bibinfo {title}
  {Hyperfine interaction in the ground state of the negatively charged nitrogen
  vacancy center in diamond},\ }\href
  {https://doi.org/10.1103/PhysRevB.79.075203} {\bibfield  {journal} {\bibinfo
  {journal} {Phys. Rev. B}\ }\textbf {\bibinfo {volume} {79}},\ \bibinfo
  {pages} {075203} (\bibinfo {year} {2009})}\BibitemShut {NoStop}%
\bibitem [{\citenamefont {Lourette}\ \emph {et~al.}(2022)\citenamefont
  {Lourette}, \citenamefont {Jarmola}, \citenamefont {Acosta}, \citenamefont
  {Birdwell}, \citenamefont {Budker}, \citenamefont {Doherty}, \citenamefont
  {Ivanov},\ and\ \citenamefont {Malinovsky}}]{lourette_temperature_2022}%
  \BibitemOpen
  \bibfield  {author} {\bibinfo {author} {\bibfnamefont {S.}~\bibnamefont
  {Lourette}}, \bibinfo {author} {\bibfnamefont {A.}~\bibnamefont {Jarmola}},
  \bibinfo {author} {\bibfnamefont {V.~M.}\ \bibnamefont {Acosta}}, \bibinfo
  {author} {\bibfnamefont {A.~G.}\ \bibnamefont {Birdwell}}, \bibinfo {author}
  {\bibfnamefont {D.}~\bibnamefont {Budker}}, \bibinfo {author} {\bibfnamefont
  {M.~W.}\ \bibnamefont {Doherty}}, \bibinfo {author} {\bibfnamefont
  {T.}~\bibnamefont {Ivanov}},\ and\ \bibinfo {author} {\bibfnamefont {V.~S.}\
  \bibnamefont {Malinovsky}},\ }\href@noop {} {\bibinfo {title} {Temperature
  {{Sensitivity}} of \$\^\{14\}\textbackslash
  mathrm\{\vphantom\}{{NV}}\vphantom\{\}\$ and \$\^\{15\}\textbackslash
  mathrm\{\vphantom\}{{NV}}\vphantom\{\}\$ {{Ground State Manifolds}}}}
  (\bibinfo {year} {2022}),\ \Eprint {https://arxiv.org/abs/2212.12169}
  {2212.12169 [cond-mat, physics:physics, physics:quant-ph]} \BibitemShut
  {NoStop}%
\bibitem [{\citenamefont {Chen}\ \emph {et~al.}(2015)\citenamefont {Chen},
  \citenamefont {Hirose},\ and\ \citenamefont
  {Cappellaro}}]{chen_measurement_2015}%
  \BibitemOpen
  \bibfield  {author} {\bibinfo {author} {\bibfnamefont {M.}~\bibnamefont
  {Chen}}, \bibinfo {author} {\bibfnamefont {M.}~\bibnamefont {Hirose}},\ and\
  \bibinfo {author} {\bibfnamefont {P.}~\bibnamefont {Cappellaro}},\ }\bibfield
   {title} {\bibinfo {title} {Measurement of transverse hyperfine interaction
  by forbidden transitions},\ }\href
  {https://doi.org/10.1103/PhysRevB.92.020101} {\bibfield  {journal} {\bibinfo
  {journal} {Phys. Rev. B}\ }\textbf {\bibinfo {volume} {92}},\ \bibinfo
  {pages} {020101} (\bibinfo {year} {2015})}\BibitemShut {NoStop}%
\bibitem [{\citenamefont {Sangtawesin}\ \emph {et~al.}(2016)\citenamefont
  {Sangtawesin}, \citenamefont {McLellan}, \citenamefont {Myers}, \citenamefont
  {Jayich}, \citenamefont {Awschalom},\ and\ \citenamefont
  {Petta}}]{sangtawesin_hyperfine-enhanced_2016}%
  \BibitemOpen
  \bibfield  {author} {\bibinfo {author} {\bibfnamefont {S.}~\bibnamefont
  {Sangtawesin}}, \bibinfo {author} {\bibfnamefont {C.~A.}\ \bibnamefont
  {McLellan}}, \bibinfo {author} {\bibfnamefont {B.~A.}\ \bibnamefont {Myers}},
  \bibinfo {author} {\bibfnamefont {A.~C.~B.}\ \bibnamefont {Jayich}}, \bibinfo
  {author} {\bibfnamefont {D.~D.}\ \bibnamefont {Awschalom}},\ and\ \bibinfo
  {author} {\bibfnamefont {J.~R.}\ \bibnamefont {Petta}},\ }\bibfield  {title}
  {\bibinfo {title} {Hyperfine-enhanced gyromagnetic ratio of a nuclear spin in
  diamond},\ }\href {https://doi.org/10.1088/1367-2630/18/8/083016} {\bibfield
  {journal} {\bibinfo  {journal} {New J. Phys.}\ }\textbf {\bibinfo {volume}
  {18}},\ \bibinfo {pages} {083016} (\bibinfo {year} {2016})}\BibitemShut
  {NoStop}%
\bibitem [{\citenamefont {Oon}\ \emph {et~al.}(2022)\citenamefont {Oon},
  \citenamefont {Tang}, \citenamefont {Hart}, \citenamefont {Olsson},
  \citenamefont {Turner}, \citenamefont {Schloss},\ and\ \citenamefont
  {Walsworth}}]{oon_ramsey_2022-1}%
  \BibitemOpen
  \bibfield  {author} {\bibinfo {author} {\bibfnamefont {J.~T.}\ \bibnamefont
  {Oon}}, \bibinfo {author} {\bibfnamefont {J.}~\bibnamefont {Tang}}, \bibinfo
  {author} {\bibfnamefont {C.~A.}\ \bibnamefont {Hart}}, \bibinfo {author}
  {\bibfnamefont {K.~S.}\ \bibnamefont {Olsson}}, \bibinfo {author}
  {\bibfnamefont {M.~J.}\ \bibnamefont {Turner}}, \bibinfo {author}
  {\bibfnamefont {J.~M.}\ \bibnamefont {Schloss}},\ and\ \bibinfo {author}
  {\bibfnamefont {R.~L.}\ \bibnamefont {Walsworth}},\ }\bibfield  {title}
  {\bibinfo {title} {Ramsey envelope modulation in {{NV}} diamond
  magnetometry},\ }\href {https://doi.org/10.1103/PhysRevB.106.054110}
  {\bibfield  {journal} {\bibinfo  {journal} {Phys. Rev. B}\ }\textbf {\bibinfo
  {volume} {106}},\ \bibinfo {pages} {054110} (\bibinfo {year}
  {2022})}\BibitemShut {NoStop}%
\bibitem [{SOM()}]{SOM}%
  \BibitemOpen
  \href@noop {} {}\bibinfo {howpublished} {See Supplemental Material for
  details.}\BibitemShut {Stop}%
\bibitem [{\citenamefont {Chen}\ \emph {et~al.}(2018)\citenamefont {Chen},
  \citenamefont {Sun}, \citenamefont {Saha}, \citenamefont {Jaskula},\ and\
  \citenamefont {Cappellaro}}]{chen_protecting_2018}%
  \BibitemOpen
  \bibfield  {author} {\bibinfo {author} {\bibfnamefont {M.}~\bibnamefont
  {Chen}}, \bibinfo {author} {\bibfnamefont {W.~K.~C.}\ \bibnamefont {Sun}},
  \bibinfo {author} {\bibfnamefont {K.}~\bibnamefont {Saha}}, \bibinfo {author}
  {\bibfnamefont {J.-C.}\ \bibnamefont {Jaskula}},\ and\ \bibinfo {author}
  {\bibfnamefont {P.}~\bibnamefont {Cappellaro}},\ }\bibfield  {title}
  {\bibinfo {title} {Protecting solid-state spins from a strongly coupled
  environment},\ }\href {https://doi.org/10.1088/1367-2630/aac542} {\bibfield
  {journal} {\bibinfo  {journal} {New J. Phys.}\ }\textbf {\bibinfo {volume}
  {20}},\ \bibinfo {pages} {063011} (\bibinfo {year} {2018})}\BibitemShut
  {NoStop}%
\bibitem [{\citenamefont {Barry}\ \emph {et~al.}(2020)\citenamefont {Barry},
  \citenamefont {Schloss}, \citenamefont {Bauch}, \citenamefont {Turner},
  \citenamefont {Hart}, \citenamefont {Pham},\ and\ \citenamefont
  {Walsworth}}]{barry_sensitivity_2020}%
  \BibitemOpen
  \bibfield  {author} {\bibinfo {author} {\bibfnamefont {J.~F.}\ \bibnamefont
  {Barry}}, \bibinfo {author} {\bibfnamefont {J.~M.}\ \bibnamefont {Schloss}},
  \bibinfo {author} {\bibfnamefont {E.}~\bibnamefont {Bauch}}, \bibinfo
  {author} {\bibfnamefont {M.~J.}\ \bibnamefont {Turner}}, \bibinfo {author}
  {\bibfnamefont {C.~A.}\ \bibnamefont {Hart}}, \bibinfo {author}
  {\bibfnamefont {L.~M.}\ \bibnamefont {Pham}},\ and\ \bibinfo {author}
  {\bibfnamefont {R.~L.}\ \bibnamefont {Walsworth}},\ }\bibfield  {title}
  {\bibinfo {title} {Sensitivity optimization for {{NV-diamond}}
  magnetometry},\ }\href {https://doi.org/10.1103/RevModPhys.92.015004}
  {\bibfield  {journal} {\bibinfo  {journal} {Rev. Mod. Phys.}\ }\textbf
  {\bibinfo {volume} {92}},\ \bibinfo {pages} {015004} (\bibinfo {year}
  {2020})},\ \Eprint {https://arxiv.org/abs/1903.08176} {arXiv:1903.08176}
  \BibitemShut {NoStop}%
\bibitem [{\citenamefont {Jacques}\ \emph {et~al.}(2009)\citenamefont
  {Jacques}, \citenamefont {Neumann}, \citenamefont {Beck}, \citenamefont
  {Markham}, \citenamefont {Twitchen}, \citenamefont {Meijer}, \citenamefont
  {Kaiser}, \citenamefont {Balasubramanian}, \citenamefont {Jelezko},\ and\
  \citenamefont {Wrachtrup}}]{jacques_dynamic_2009}%
  \BibitemOpen
  \bibfield  {author} {\bibinfo {author} {\bibfnamefont {V.}~\bibnamefont
  {Jacques}}, \bibinfo {author} {\bibfnamefont {P.}~\bibnamefont {Neumann}},
  \bibinfo {author} {\bibfnamefont {J.}~\bibnamefont {Beck}}, \bibinfo {author}
  {\bibfnamefont {M.}~\bibnamefont {Markham}}, \bibinfo {author} {\bibfnamefont
  {D.}~\bibnamefont {Twitchen}}, \bibinfo {author} {\bibfnamefont
  {J.}~\bibnamefont {Meijer}}, \bibinfo {author} {\bibfnamefont
  {F.}~\bibnamefont {Kaiser}}, \bibinfo {author} {\bibfnamefont
  {G.}~\bibnamefont {Balasubramanian}}, \bibinfo {author} {\bibfnamefont
  {F.}~\bibnamefont {Jelezko}},\ and\ \bibinfo {author} {\bibfnamefont
  {J.}~\bibnamefont {Wrachtrup}},\ }\bibfield  {title} {\bibinfo {title}
  {Dynamic {{Polarization}} of {{Single Nuclear Spins}} by {{Optical Pumping}}
  of {{Nitrogen-Vacancy Color Centers}} in {{Diamond}} at {{Room
  Temperature}}},\ }\href {https://doi.org/10.1103/PhysRevLett.102.057403}
  {\bibfield  {journal} {\bibinfo  {journal} {Phys. Rev. Lett.}\ }\textbf
  {\bibinfo {volume} {102}},\ \bibinfo {pages} {057403} (\bibinfo {year}
  {2009})}\BibitemShut {NoStop}%
\bibitem [{\citenamefont {Kornack}\ \emph {et~al.}(2005)\citenamefont
  {Kornack}, \citenamefont {Ghosh},\ and\ \citenamefont
  {Romalis}}]{kornack_nuclear_2005}%
  \BibitemOpen
  \bibfield  {author} {\bibinfo {author} {\bibfnamefont {T.~W.}\ \bibnamefont
  {Kornack}}, \bibinfo {author} {\bibfnamefont {R.~K.}\ \bibnamefont {Ghosh}},\
  and\ \bibinfo {author} {\bibfnamefont {M.~V.}\ \bibnamefont {Romalis}},\
  }\bibfield  {title} {\bibinfo {title} {Nuclear {{Spin Gyroscope Based}} on an
  {{Atomic Comagnetometer}}},\ }\href
  {https://doi.org/10.1103/PhysRevLett.95.230801} {\bibfield  {journal}
  {\bibinfo  {journal} {Phys. Rev. Lett.}\ }\textbf {\bibinfo {volume} {95}},\
  \bibinfo {pages} {230801} (\bibinfo {year} {2005})}\BibitemShut {NoStop}%
\end{thebibliography}%

\newpage
\clearpage
\setcounter{section}{0}
\setcounter{equation}{0}
\setcounter{figure}{0}
\setcounter{table}{0}
\setcounter{page}{1}
\makeatletter
\renewcommand{\theequation}{S\arabic{equation}}
\renewcommand{\thesection}{S\arabic{section}}
\renewcommand{\thefigure}{S\arabic{figure}}


\title{Supplementary Materials}
\maketitle
\begin{widetext}
\section{Derivation of Enhancement Factor}
Here, we derive the exact expression for the transverse Zeeman coupling enhancement factor of the nuclear spin for the $^{15}$NV center system for $m_S =  0$ \cite{chen_measurement_2015, sangtawesin_hyperfine-enhanced_2016}.  The ground state Hamiltonian is given by Eq. (\ref{eq:Hamiltonian}). Under an applied magnetic field $B_z$, the Hamiltonian can be decomposed into secular $H_{||}$ and non-secular terms $H_{\perp}$:
\begin{equation}
    H_{||} = DS_z^2 + \gamma_e B_z S_z + \gamma_n B_z I_z + A_{zz}S_zI_z
\end{equation}
\begin{equation}
    H_{\perp} = \frac{A_{\perp}}{2}(S_+I_- + S_-I_+).
\end{equation}
where $S_{\pm} = S_x \pm i S_y$ and $I_{\pm} = I_x \pm i I_y$. The total Hamiltonian can then be diagonalized by rotating the zero quantum (ZQ) subspaces with $U_{ZQ} = e^{-i(\theta^-\sigma_y^- + \theta^+\sigma_y^+)}$ where $\sigma_y^+ = i(|+1, -\frac{1}{2}\rangle \langle 0, +\frac{1}{2}| - |0, +\frac{1}{2}\rangle \langle +1, -\frac{1}{2}|)$ and $\sigma_y^- = i(|0, -\frac{1}{2}\rangle \langle -1, +\frac{1}{2}| - |-1, +\frac{1}{2}\rangle \langle 0, -\frac{1}{2}|)$ and \begin{equation}
    \tan(2 \theta^+) = \frac{2A_{\perp}}{D + \gamma_eB_z - \gamma_NB_z-A_{zz}/2}
\end{equation}
\begin{equation}
    \tan(2 \theta^-) = \frac{-2A_{\perp}}{D - \gamma_eB_z + \gamma_NB_z - A_{zz}/2}
\end{equation}

Applying an additional transverse magnetic field in the NV frame (for simplicity and in context with our paper, consider only $B_x$, but a magnetic field component $B_y$ would see an equivalent enhancement) introduces an interaction Hamiltonian $H_x = B_x (\gamma_e S_x + \gamma_N I_x)$.  Under the unitary transformation, $\hat{H_x} = U_{ZQ}H_xU_{ZQ}^{\dagger}$ and    keeping only the nuclear spin terms, we obtain the effective Hamiltonian for the nuclear spin (where $\alpha_{m_s}$ is the enhancement factor for the electronic spin state $m_s$)
\begin{equation}
    \hat{H_{x}}^I = \gamma_N(\alpha_{+1}|+1\rangle \langle +1| + \alpha_{0}|0\rangle \langle 0| + \alpha_{-1}|-1\rangle \langle -1|)
\end{equation}
with 
\begin{gather*}
    \alpha_{0} = \cos(\theta^+)\cos(\theta^-)-  \frac{\gamma_e}{\gamma_n}\sin(\theta^+ - \theta^-) 
\end{gather*}
\begin{figure}[h]
\includegraphics[width=0.5\textwidth]{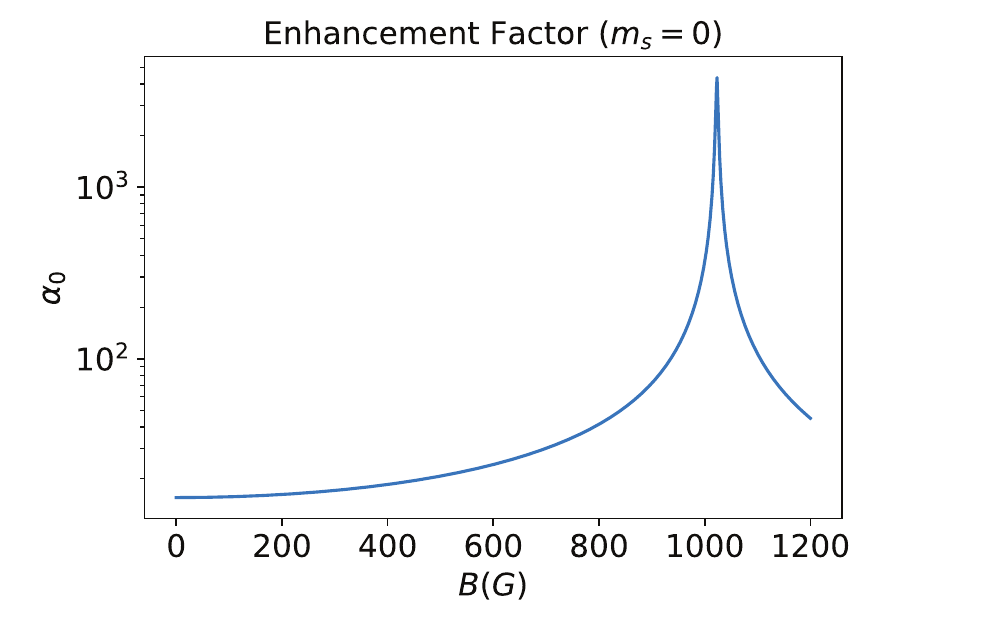}
\caption{\label{fig_supp:enhancement}
Predicted enhancement factor $|\alpha_0|$ as a function of $B_z$. }
\end{figure}
Thus, for an applied magnetic field $\Vec{B} = (B_x, 0, B_z)$ on the NV center, we obtain the effective nuclear spin Hamiltonian as presented in Eq. (\ref{eq:H_I}).  For small fields $\alpha_0 \approx 15.5$ and reaches a finite maximum value near GSLAC with $\alpha_0 \approx \frac{\gamma_e}{\sqrt{2}\gamma_n}$ shown in Fig. \ref{fig_supp:enhancement}.

\label{thesection:enhancement factor}
\section{Protocol Details}

\subsection{Nuclear Polarization and Readout}
There are several methods to polarize the nuclear spin intrinsic to the NV center: we can optically pump the nuclear spin by setting the static magnetic field close to the ground or excited level crossing, or we can use a sequence of selective microwave and rf pulses to transfer polarization from the NV (Fig. \ref{fig:pol_sequence}(a).)  Under misalignment of the NV spin, however, optical polarization of the nuclear spin is significantly suppressed \cite{jacques_dynamic_2009}.  Thus to achieve efficient polarization of the nuclear spin for our gyroscope sensing protocol, especially in the case of initialization during a rotation (which causes NV misalignment), it is preferable to polarize the nuclear spin to  $|m_I = +\frac{1}{2}\rangle$ using a polarization sequence that coherently transfers the electron polarization to the nuclear spin.  

Working in the $m_S =  0, -1$ states, we first transfer the population from the $|m_S =  0, m_I=-\frac12\rangle$ spin state to $|m_S =  -1, m_I=-\frac12\rangle$ using a selective microwave (mw) $\pi$-pulse with a Rabi frequency smaller than the hyperfine splitting strength.  We then apply a selective rf $\pi$-pulse that transfers the $|m_S =  -1, m_I = -\frac{1}{2}\rangle$ population to $|m_S =  -1, m_I = +\frac{1}{2}\rangle$.  Finally, a green laser pulse is applied to transfer the population back to the $|m_S =  0\rangle$ manifold while preserving the nuclear spin state. The system is ultimately polarized to $|m_S =  0, m_I = +\frac{1}{2} \rangle$ as an initial state of the rotation sensing protocol.  

Optical readout of the nuclear spin state is accomplished with mw mapping pulses at the end of the pulse sequence, as shown in Fig.~\ref{fig:pol_sequence}(a).  

To achieve efficient initialization and readout, we  prefer the NV to be closely aligned with the magnetic field when performing the initialization and readout operations. 
\begin{figure*}
\includegraphics[width=\textwidth]{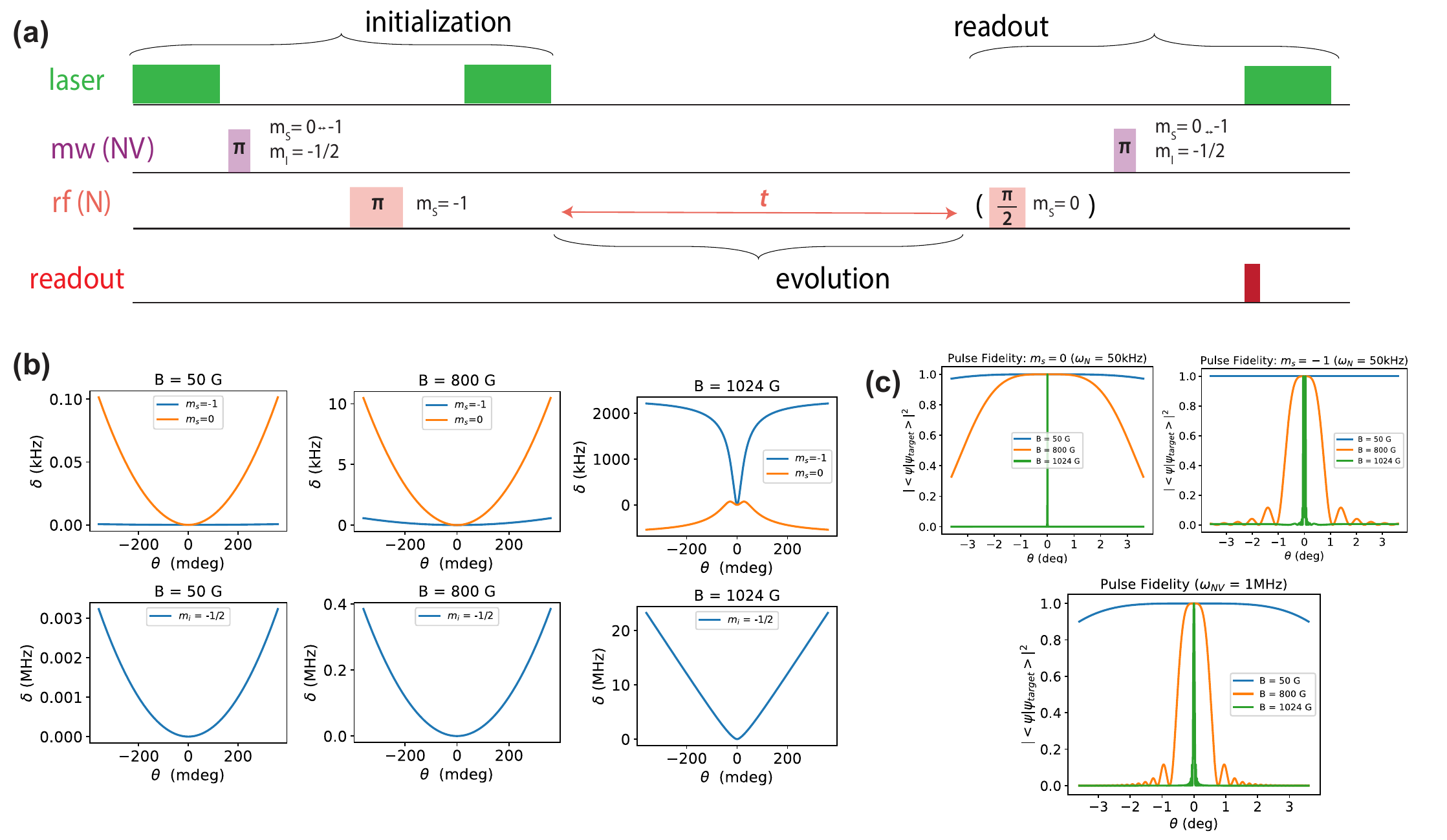}
\caption{\label{fig:pol_sequence} \textbf{Example initialization and readout protocol} (a) Basic sensing pulse sequence.  The nuclear spin is dynamically polarized using selective microwave and rf pulses.  After a sensing time $t$, the nuclear spin is mapped onto the NV and overlap is read out to extract the rotation rate.  (b) NV electronic and nuclear spin transition frequency shifts (from alignment at $\theta = n \pi$) as a function of magnetic field misalignment $\theta$ at different magnetic field strengths. The top three panels are transition frequencies between the two nuclear spin levels under different magnetic fields. The bottom three panels are transition frequencies between the two electronic spin states.   (c) These frequency shifts are used to calculate the fidelity of the nuclear and electronic spin (selective on the nuclear spin state $m_I=\pm 1/2$) $\pi$ pulses. The Rabi frequency for the nuclear spin used here 50~kHz and the electronic spin is 1~MHz.  This allows the assume pulse duration of $t_{\pi} \sim 0.5 \mu$s used to compute the pulse fidelity factor in Fig.3(c) in the main text.} 
\end{figure*}
Indeed,  the transition frequencies of the NV electronic and nuclear spins change during the rotation due to their misalignment with the magnetic field in the NV frame.  Thus, when applying a pulse, these energy shifts can affect the pulse fidelity for the polarization and readout sequences.  
Because this change is larger at higher magnetic fields (Fig.~\ref{fig:pol_sequence}(b)), even though the enhancement factor is larger, the sensitivity may be practically limited at large magnetic fields by errors in initialization and qubit manipulation.  We note that this challenge could be addressed by setting optimal pulse timing settings for the polarization/readout sequence in an adaptive way, as control optimization requires an initial estimation of the rotation rate or an independent measure of the resonance frequency by interleaving the gyroscope sequence with nuclear Ramsey or NV ODMR experiments.
For simplicity, here we discuss the worst-case scenario, where we assume to just use a single frequency for the microwave and rf pulses.   As relevant to quantum sensing, We consider the state fidelity for 
an initial state $|\psi_i\rangle$ and a target state $\psi_{\text{target}}$, 
$\mathcal{F} = \abs{\langle \psi_{\text{target}}|U|\psi_i \rangle}^2$.
We can then evaluate the effect  of the detuning induced by a $\theta$-misalignment of the field on a $\pi$-pulse $U_{\pi}(t)$. For $\ket{\psi_i}=\ket{0}$ and $\ket{\psi_{\text{target}}}=\ket{1}$, 
the $\pi$-pulse fidelity with detuning $\delta$ and amplitude $\Omega_d$ is given by $\mathcal{F}(\Omega_d, \delta) = \frac{\Omega_d^2}{\Omega_R^2}\sin^2(\frac{\Omega_R t}{2})$, 
with  $\Omega_R = \sqrt{\Omega_d^2+\delta^2}$ the effective Rabi frequency  and $t = \pi/\Omega_d$ the pulse duration.  As shown in Fig.~\ref{fig:pol_sequence}(b), when the magnetic field is not close to the GSLAC condition, the nuclear spin transition frequencies are relatively stable against a small field misalignment, and the change of the electron spin transition frequency is smaller than the typical Rabi frequency $\Omega\sim1$~MHz. As a result, high-fidelity control can be achieved even under the rotation as long as the Rabi frequency is larger than the transition frequency change during the rotation and smaller than the difference $\omega_n^{m_S =  -1} - \omega_n^{m_S =  -0} \approx 3$MHz, as shown in Fig.~\ref{fig:pol_sequence}(c).
In the main text  we accounted for these potential errors in qubit manipulation when estimating the sensitivity of the protocol (Fig~3). In the sensitivity formula, we introduced  a degraded contrast of the signal: $C \ge \mathcal{F}C_0$, where $\mathcal{F}(B)$ is the B-field dependent pulse fidelity  and  $C_0 \sim 2\%$ the contrast for a typical NV ensemble. We note that the imperfect selectivity of the readout pulse can also lead to additional contrast degradation. These imperfections can be improved with various optimal control strategies such as narrow-band and broad-band pulse designs.

\begin{figure*}[htbp]
\includegraphics[width=\textwidth]{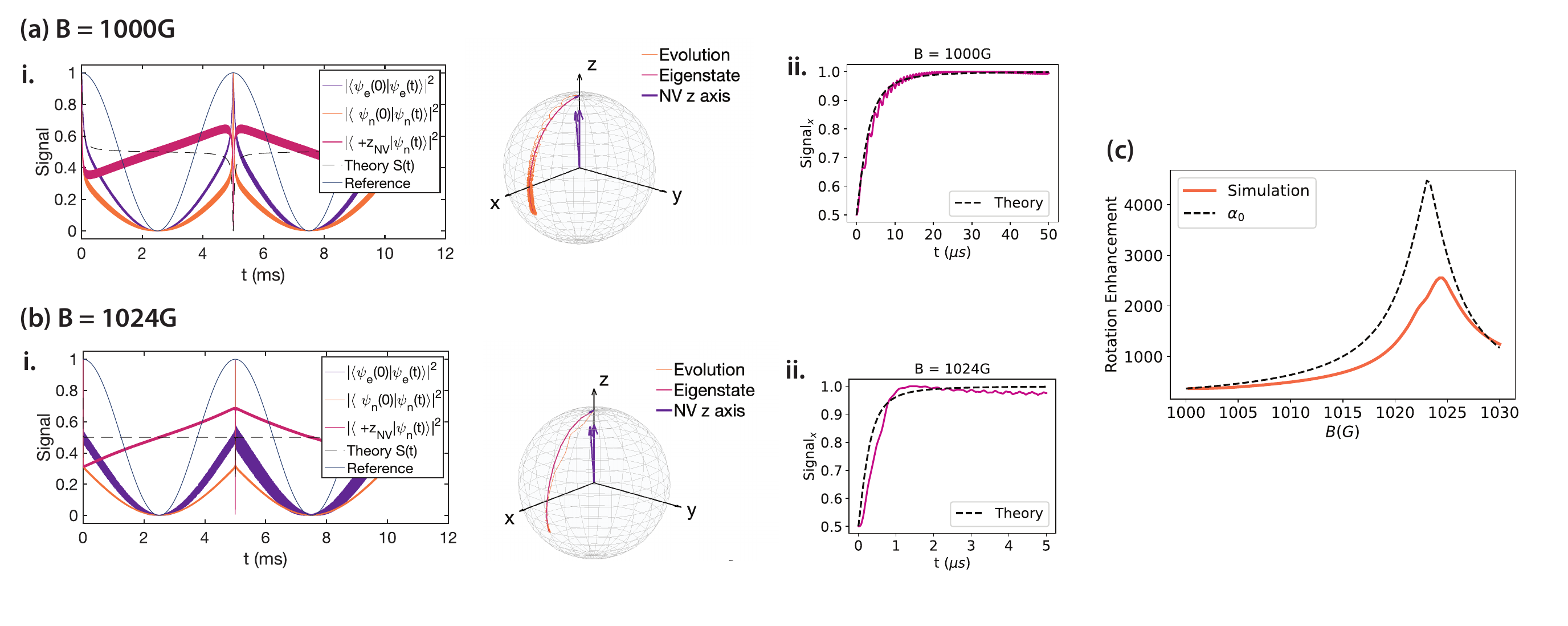}
\caption{\label{fig_supp:GSLAC} \textbf{Hyperfine-enhanced gyroscope at large magnetic fields} (a) i. Electro-nuclear spin system evolution during a ($2\pi$)0.1kHz rotation at B = 1000 G.  The left plot shows the measured nuclear spin population along the $\hat{z}_{\text{NV}}$ axis and the evolution of the electronic and nuclear spin states, compared to the reference diamond rotation and the predicted signal from \ref{eq:S(t)}. An accompanying schematic of the NV and nuclear spin evolution on the Bloch sphere is shown on the right.  ii. Simulated measurement signal of the nuclear spin population along $\hat{x}_{\text{NV}}$ axis, used to extract the rotation rate. (b) System evolution at GSLAC (B = 1024 G). (c) Simulated rotation angle enhancement factor for $B \ge 1000$G, compared to analytically-derived enhancement factor $|\alpha_0|$.}
\end{figure*}

\subsection{Performance near the GSLAC}
While the enhancement factor $\alpha_0$  reaches its maximum value $\approx \frac{\gamma_e}{\sqrt{2} \gamma_n} \sim 4 \times 10^3$ near the GSLAC ($B \approx D/\gamma_e \approx 1024$G), in that scenario  the behavior of the gyroscope becomes more complex. Due to its small energy gap, the electron spin not only is  strongly  mixed with the nuclear spin but it also no longer follows a completely adiabatic evolution.  Nevertheless, simulation results indicate that even at large magnetic fields, the enhancement of the nuclear spin rotation compared to the electron spin is  maintained. 
The hyperfine interaction still yields a relative rotation between the two spins, and the rotation rate can still be recovered (Fig. \ref{fig_supp:GSLAC}i). While the observed enhancement factor near the GSLAC is lower than the analytical prediction by nearly a factor of 2, it could still provide gain a three-order of magnitude sensitivity enhancement.  As mentioned above, this enhancement might be reduced due to control limitations, although adaptive methods will mitigate those issues. 

\subsection{Gyroscope Sensitivity}
\subsubsection{Hyperfine-Enhanced Gyroscope}
The sensitivity of the gyroscope for an ensemble of $N$ spins is given by 
\begin{equation}
    \eta = \frac{\sigma e^{t/\tau}\sqrt{t+t_d}}{2 C \sqrt{N} \abs{\partial_\Omega S(t)}}
\end{equation}
where $t$ is the total sensing time, $t_d$ is the deadtime, $\tau$ is the coherence time, $C$ is the readout efficiency parameter that is typically set by the signal contrast and photon collection efficiency, $N$ is the total number of spins, and $S(t)$ and $\sigma$ its standard deviation from the spin projection noise. 

In practice, to optimize the sensitivity, we can maximize the signal slope by measuring the nuclear spin population along the $\hat{x}_{\text{NV}}$ (which can be accomplished by applying a $\pi/2$-pulse to the nuclear spin before readout) at the time of near-alignment condition, yielding the signal:
\begin{equation}
    S_x(t) = \frac{1}{2}\left( 1 + \frac{\alpha_0 \sin(\Omega t)}{\sqrt{\alpha_0^2 \sin^2(\Omega t) + \cos^2(\Omega t)}}\right).
\end{equation}
To ensure continuous operation of the protocol as well as efficient polarization and readout, we require $t+t_d \approx \ceil{\frac{\Omega t}{\pi}}(\frac{\pi}{\Omega})$ and assume the initialization and readout times ($t_d\sim 3 \mu$s under large laser power and fast control pulses) are small in comparison to period of the oscillation $2\pi/\Omega$. The sensitivity then gives
\begin{equation}
\eta(\Omega \ll \gamma_nB, B, t) = \frac{1}{(\frac{\alpha_0}{2})\abs{\frac{\cos(\Omega t)}{(\alpha_0^2 \sin^2(\Omega t) + \cos(\Omega t)^2)^{3/2}}}} \left( \frac{e^{t/\tau}\sqrt{\ceil{\frac{\Omega t}{\pi}}(\frac{\pi}{\Omega})}}{2 C \sqrt{N} t } \right)\geq  \frac{e^{t/\tau}}{ C \sqrt{N} \alpha_0 \sqrt{t} }
\label{eq:sensitivity_supp}
\end{equation}

Hence the optimal sensitivity is achieved when the sensing time $ t \approx k\pi/\Omega$ and the sensitivity has an enhancement of $\alpha_0$ in comparison to the conventional scheme based on nuclear Ramsey (where $\abs{\partial_{\Omega}S(t)}_{\text{max}} = t/2$).   The ultimate bound on the sensitivity occurs near GSLAC, where  $\alpha_0 \approx \frac{\gamma_e}{\sqrt{2} \gamma_n}$, and with sensing time $t \approx \tau/2$.  

Assuming a typical NV ensemble diamond sample chip of volume $V = 1\text{ mm}^3$ with $N = 2.5\times 10^{14}$ spin sensors,  detection efficiency $C \sim 2\%$ and coherence time $\tau = 1.5T_{1e} \approx 7.5$ms, the sensitivity bound (Eq.~\eqref{eq:sensitivity_supp}) can 
achieve $\eta(\Omega \ll \gamma_nB, B \approx 1024\text{G}, t \approx \tau/2) \approx 1 \times 10^{-3}$ (mdeg/s)/$\sqrt{\text{Hz}}$, comparable with atomic gyroscopes~\cite{kornack_nuclear_2005}.  

{Nevertheless, for slow rotations such that the signal decay time is on the same order as the rotation period when $\tau \sim \pi/\Omega$, there is a tradeoff: the measurement needs to be made as quickly as possible after initialization of the electron spin to maintain the high sensitivity enhancement factor in the signal,  which has a conflict with maximizing the sensing time.  In this case, the sensitivity enhancement is dictated by the factor $\partial_{t}S(t)$.  We note that for even lower rates where $\Omega \tau \ll 1$, the sensing time is again limited by the coherence time and the deadtime, thus reaching the sensitivity bounds of the case where $\tau \gg \pi/\Omega$.  }


\subsubsection{Inertial Gyroscope}

When sensing fast rotations $\Omega\gg \gamma_n B$, the electron spin and nuclear spin still need to be nearly aligned with the magnetic field when reading out, so we still require $t+t_d \approx \ceil{\frac{\Omega t}{\pi}}(\frac{\pi}{\Omega})$ and the sensitivity  (measured along $\hat{x}_{\text{NV}}$) is given by
\begin{equation}
\label{eq_supp: fast_sensitivity}
    \eta(\Omega \gg \gamma_nB, B, t) = \frac{1}{(\frac{1}{2})\abs{\cos(\Omega t)}} \left( \frac{e^{t/\tau}\sqrt{\ceil{\frac{\Omega t}{\pi}}(\frac{\pi}{\Omega})}}{2 C \sqrt{N} t } \right)\geq\frac{e^{t/\tau}}{ C \sqrt{N} \sqrt{t} }
\end{equation}
The lower bound for the optimal sensitivity of the sensor can thus be approximated $\eta_{\text{opt}} \approx \frac{\sqrt{2e}}{C \sqrt{N\tau}} \approx 4.91$ (mdeg/s)$/\sqrt{\text{Hz}}$ for $\tau \approx 7.5$ms. 

\subsubsection{Comparison with other protocols}
The most conventional method in sensing a rotation is using the nuclear Ramsey.  The Ramsey protocol using the nuclear spin has a signal of the form $S(t) = \frac{1}{2}(1 + \sin(\Omega  t))$.
Thus, the sensitivity is similar to that given in Eq.~\ref{eq_supp: fast_sensitivity}:
\begin{equation}
    \eta_{\text{N-Ramsey}}(t)= \frac{1}{(\frac{1}{2})\abs{\cos(\Omega t)}}\left( \frac{e^{t/\tau}\sqrt{t + t_d}}{2 C \sqrt{N} t } \right)
\end{equation}
and yields a lower bound to the sensitivity $\eta_{\text{opt}} \approx \frac{\sqrt{2e}}{C \sqrt{N\tau}}$.  The coherence time for the Ramsey sequence is limited by the spin dephasing time $T_{2n}^* \leq 1.5T_{1e}$~\cite{chen_protecting_2018}.  Thus, even in the fast rotation sensing regime (without hyperfine-enhancement) of the gyroscope developed in this work,  the sensitivity is slightly improved compared to the conventional nuclear spin Ramsey scheme, as we found $\tau > 1.5T_{1e}$ in numerical simulations. 

Alternatively, with the assistance of the external magnetic field along $\hat{z}$ (like in the setup discussed in our work), the change in the electron spin transition frequency can also be used to extract the rotation rate (for a transverse rotation along $\hat{y}$).  In this case, similar to conventional magnetometry, this is tested with a Ramsey experiment with the electron spin, ultimately limited by the spin dephasing time $\tau = T_{2e}^* \ll T_{1e}$ ~\cite{wang_emulated_2023}. The signal of the Ramsey is given by $S(t) = \frac{1}{2}(1 + \sin(\omega_e t))$, where $\omega_e = \omega_e(\theta)$ is the rotation-angle dependent electron spin transition frequency.  Thus, the sensitivity is given by 
\begin{equation}
    \eta_{\text{NV-Ramsey}}(t) = \frac{1}{(\frac{1}{2})\abs{\cos(\omega_e t) \frac{\partial \omega_e}{\partial \theta}}}\left( \frac{e^{t/\tau}\sqrt{t+t_d}}{2 C \sqrt{N} t^2 } \right).
\end{equation}
In an electronic spin ensemble, the spin dephasing time is typically 2 to 3 order-of-magnitude shorter than in a nuclear spin ensemble, while the derivative $\frac{\partial \omega_e}{\partial \theta}$ increases with an increase in the magnetic field and can improve the sensitivity. Thus it is not straightforward to analytically compare this scheme to other ones and here we discuss a few special conditions.  Near GSLAC at $B \approx 1000$G, $\abs{\frac{\partial \omega_e}{\partial \theta}}_{\text{max}} \approx (2\pi) 4\times10^3$ MHz/rad, and although $T_{2e}^* \ll T_{1e}$, the sensitivity is improved compared to the nuclear spin Ramsey, with $\eta_{\text{opt}} \approx 0.2 $ (mdeg/s)$/\sqrt{\text{Hz}}$ for $T_{2e}^* = 0.7\mu s$, still worse than the hyperfine-enhanced gyroscope.


A sensitivity comparison of all conventional NV-based gyroscopes is shown in the main text Fig.~2e.

\subsection{Adiabatic Range for Hyperfine-Enhanced Sensing}
The hyperfine-enhanced gyroscope protocol requires that the nuclear spin quantization axis follow the effective magnetic field axis adiabatically. Because we assume that the gyroscope operates in regimes where the rotation rate is expected to be significantly less than the ZFS $D = 2\pi \times 2.87$ GHz, the electron spin is expected to follow the eigenstate set by the crystalline axis of the diamond.  Nevertheless, because the $^{15}$N does not have a large quadrupole term, the energy splitting of the nuclear spin is dependent on the magnetic field strength, and the rotation has to be sufficiently slow with respect to $\gamma_nB$ such that the evolution of the nuclear spin remains adiabatic and the hyperfine-enhanced signal can be measured.  Thus, to provide a quantifiable range for the gyroscope, we derive the adiabatic criteria for the nuclear spin under the rotation. 

The effective Hamiltonian for the nuclear spin under the rotation of rate $\Omega$ (in the NV-frame) is given by: 
\begin{equation}
    H(t) = \gamma_n(B_z(t)I_z + \alpha_0B_x(t)I_x) - \Omega I_y
\end{equation}
where $B_z(t) = B\cos(\Omega t)$ and $B_x(t) = -B\sin(\Omega t)$.  The time-dependent Hamiltonian has two orthogonal eigenstates $|\psi_{\pm}(t) \rangle$ such that such that $H(t)|\psi_{\pm}(t)\rangle = E_{\pm}|\psi_{\pm}(t)\rangle$.
The time evolution operator is given by:
\begin{align}
    U(t) = \mathcal{T}\exp(-i \int_0^t H(t^\prime) \delta t^\prime) = \exp(-i \sum_{\phi = \psi_\pm} \int_0^t E_\phi (t') |\phi \rangle \langle \phi| dt') =  \sum_{\phi} \exp(-i \int_0^t E_{\phi}(t') dt') |\phi \rangle \langle \phi |,
\end{align}
where $\mathcal{T}$ is the time-ordering operator.
Thus, we can write the time-dependent wavefunction:
\begin{equation}
    |\psi(t)\rangle = \sum_{\phi = \psi_\pm} c_{\phi}(t)\exp(-i \int_0^t E_{\phi}(t') dt') |\phi \rangle
\end{equation}
for coefficients $c_{\phi}$.  In our protocol, we initiate the nuclear spin in the $|+\frac{1}{2}\rangle$ state, so let $c_+(0) = 1, c_-(0) = 0$. From the time-dependent Schrodinger equation, we obtain:
\begin{gather}
    i \frac{\partial |\psi(t)\rangle}{\partial t} = H(t)|\psi(t)\rangle \\
    i\sum_{\phi} \exp(-i \int_0^t E_{\phi}(t') dt') \left[ \frac{d c_{\phi}}{dt} |\phi \rangle + c_{\phi} \frac{\partial |\phi \rangle}{\partial t} - iE_{\phi} c_{\phi} |\phi \rangle \right] = \sum_{\phi} c_{\phi} E_{\phi} \exp(-i \int_0^t E_{\phi}(t') dt') |\phi \rangle 
\end{gather}
We multiply both sides with the state $\langle \phi' \ne \phi |$ and divide by $i \exp(-i \int_0^t E_{\phi}(t') dt')$:
\begin{gather}
    c_{\phi} \langle \phi' | \frac{\partial \phi}{\partial t} \rangle + \exp(-i \int_0^t (E_{\phi'} - E_{\phi})(t') dt')\left[ \frac{d c_{\phi'}}{dt} + c_{\phi'} \langle \phi' | \frac{\partial \phi'}{\partial t} \rangle \right] = 0
\end{gather}
thus giving an expression for the coefficient $c_{\phi'}$:
\begin{gather}
\label{eq: adiabatic_eq}
    \frac{d c_{\phi'}}{dt} = -c_{\phi'} \langle \phi' | \frac{\partial \phi'}{\partial t} \rangle + \exp(i \int_0^t (E_{\phi'} - E_{\phi})(t') dt') c_{\phi} \langle \phi' | \frac{\partial \phi }{\partial t} \rangle 
\end{gather}
The adiabatic approximation requires that the state remains in its eigenstate and evolves independently of other states and thus $\langle \phi' | \frac{\partial \phi'}{\partial t} \rangle \gg \langle \phi' | \frac{\partial \phi }{\partial t} \rangle $.  
For the adiabatic part, we have the evolution 
\begin{gather}
    \frac{d c_{\phi'}}{dt} = -c_{\phi'} \langle \phi' | \frac{\partial \phi'}{\partial t} \rangle \\
    \frac{\partial \phi'}{\partial t} = -i H(t) |\phi'\rangle = -i E_{\phi'}(t) | \phi' \rangle \\
    c_{\phi'}(t) = c_{\phi'}(0) \exp(-\int_0^t \langle \phi'(t')|\frac{\partial \phi'(t')}{\partial t} \rangle dt') = c_{\phi'}(0) \exp(i\int_0^t E_{\phi'}(t') dt')
\end{gather}
Thus, in the adiabatic approximation of the evolution, the population in the states remain the same, while they acquire a (Berry) phase.  The second term in Eq. ($\ref{eq: adiabatic_eq}$) describes the nonadiabatic effects of the evolution and the nonadiabatic coupling $\langle \phi' | \frac{\partial \phi }{\partial t} \rangle$ determines the magnitude of effects of non-adiabaticity.   
We see that 
\begin{gather}
    \langle \phi' | \left[\frac{\partial}{\partial t}(H |\phi \rangle) = \frac{\partial}{\partial t}H |\phi \rangle + H |\frac{\partial \phi}{\partial t}\rangle =  \frac{\partial E_{\phi}}{\partial t}|\phi \rangle +E_{\phi} \frac{\partial |\phi \rangle }{\partial t}\right] \Rightarrow 
    \langle \phi' | \frac{\partial H }{\partial t} |\phi \rangle + E_{\phi'}\langle \phi' | \frac{\partial \phi }{\partial t} \rangle = E_{\phi}\langle \phi' | \frac{\partial \phi }{\partial t} \rangle
\end{gather}

and establish an adiabatic criteria for eigenstates $|\phi \rangle, |\phi' \rangle = |\psi_\pm \rangle$ :
\begin{equation}
\label{eq_supp: adiabatic criteria}
    \abs{\langle \phi' | \frac{\partial \phi }{\partial t} \rangle}^2 = \frac{\abs{\langle \phi' | \frac{\partial H }{\partial t} |\phi \rangle}^2}{(E_{\phi} - E_{\phi'})^2} \ll E_{\phi'}^2
\end{equation}

Explicitly, 
\begin{gather}
    H(t) = \frac{1}{2}B\gamma_n\left[\begin{matrix}\cos{\left(\Omega t \right)} & - \alpha_0 \sin{\left(\Omega t \right)} + i \Omega\\-  \alpha_0\sin{\left(\Omega t \right)} - i \Omega & - \cos{\left(\Omega t \right)}\end{matrix}\right] \\
    \frac{\partial H(t)}{dt} = \frac{1}{2}B\gamma_n\left[\begin{matrix}- \Omega \sin{\left(\Omega t \right)} & - \alpha_0\Omega \cos{\left(\Omega t \right)}\\- \alpha_0 \Omega \cos{\left(\Omega t \right)} & \Omega \sin{\left(\Omega t \right)}\end{matrix}\right]
\end{gather}
Solving for the instantaneous eigenstates and eigenenergies of $H(t)$, we obtain:
\begin{gather}
    |\psi_{\mp}(t)\rangle = B\gamma_n\left[\begin{matrix}- \frac{- \alpha_0\sin{\left(\Omega t \right)} + i \Omega}{\cos{\left(\Omega t \right)} \pm \sqrt{ \alpha_0^{2}\sin^{2}{\left(\Omega t \right)} + \cos^{2}{\left(\Omega t \right)} + (\frac{\Omega}{B\gamma_n})^{2}}}\\1\end{matrix}\right] \\
    E_{\mp} = \mp \frac{B\gamma_n}{2}\sqrt{ \alpha_0^{2}  \sin^{2}{\left(\Omega t \right)} +  \cos^{2}{\left(\Omega t \right)} + (\frac{\Omega}{B\gamma_n})^{2}}
\end{gather}

From Eq.~\eqref{eq_supp: adiabatic criteria}, the nuclear spin remains adiabatic when 
\begin{equation}
    \mathcal{M}(t) = \frac{\abs{\langle \psi_+| \frac{\partial \psi_- }{\partial t} \rangle}^2}{E_+^2} \ll 1
\end{equation}

Thus, to maintain the adiabaticity of the nuclear spin during the evolution, we require $\mathcal{M}(t) \ll 1$ during the rotation.  By numerically computing $\mathcal{M}(t)$ for different values of $B$ over the course of $\theta = \Omega t \in [0, \pi]$ for a chosen $\Omega = 0.1$kHz shown in Fig.~\ref{fig_supp:adiabaticity}, we see that $\mathcal{M}(t)$ reaches its minimum value at $\theta = \pi/2$, where the effective magnetic field is the largest. Calculating $\mathcal{M}(t)$ at a time $t = t_0$ allows us to define a cutoff $\epsilon$ such that the system is sufficiently adiabatic to achieve a measurable signal of the rotation. {$\epsilon$ is defined as the maximal acceptable signal uncertainty $\langle \Delta S \rangle$ (as shown in main text Fig.~2c), which then sets a standard to quantify the projected range of frequencies detectable by the hyperfine-enhanced gyroscope for a desired precision. }  
\begin{figure*}[htbp]
\includegraphics[width=0.5\textwidth]{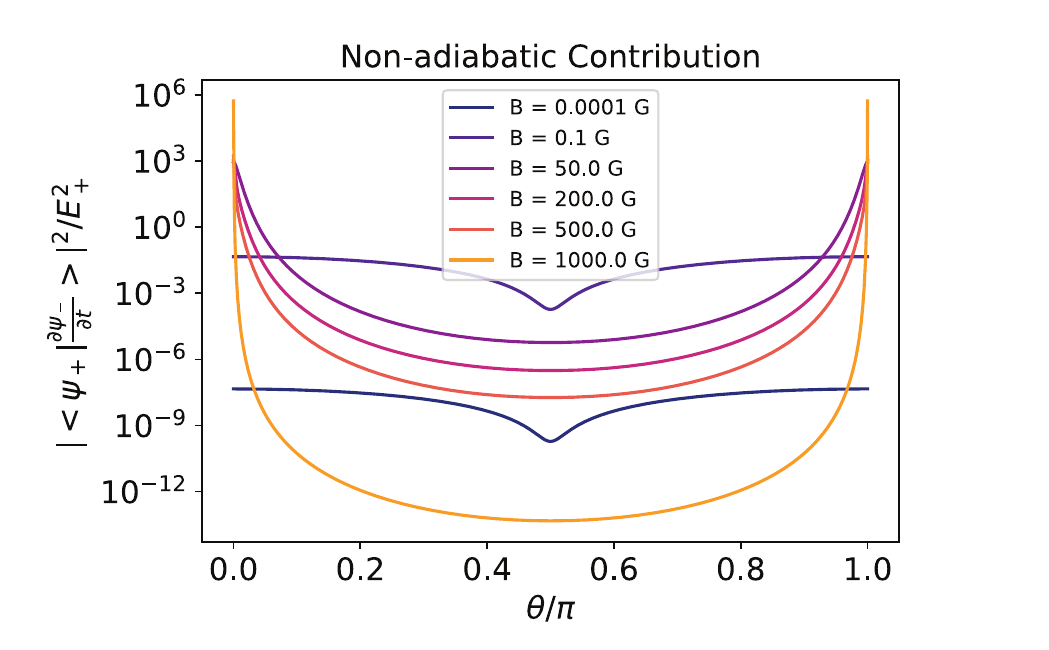}
\caption{\label{fig_supp:adiabaticity} \textbf{Nuclear Spin Eigenstate Deviation} The eigenstate deviation $\mathcal{M}(t)$ as a function of the rotation angle (at a fixed rotation rate $(2\pi) 0.1$kHz), for various values of magnetic field strength.  As $B$ increases, $\mathcal{M}(t)$ is suppressed and the system becomes adiabatic and we can perform sensing in the hyperfine-enhanced regime.  As $B$ decreases, $\mathcal{M}(t)$ increases, and the system is in the intermediate regime where a rotation signal is difficult to extract due to the non-adiabatic evolution of the nuclear spin.  Nevertheless, for a sufficiently small $B$,  $\mathcal{M}(t)$ begins to decrease again and we can sense in the inertial regime where the nuclear spin is fixed in its initial eigenstate during the rotation.}
\end{figure*}

At $\theta = \pi/2$, we can simplify 
\begin{equation}
    \mathcal{M}_{\text{min}} = \frac{4 (\frac{\Omega}{B\gamma_n})^2}{(\alpha_0^2 + (\frac{\Omega}{B\gamma_n})^2)^2}
\end{equation}
If we require that the adiabatic condition for the nuclear spin satisfy $\mathcal{M}_{\text{min}} < \xi^2$ for some $\xi$, and assume that $\frac{\Omega}{B\gamma_n} \ll 1$, to first order, we can approximate:
\begin{equation}
    \Omega < \xi \alpha_0^2 B\gamma_n
\end{equation}
Thus, the maximum detectable rotation rate by the hyperfine-enhanced gyroscope is given by $\Omega_{\text{max}} \propto \alpha_0^2 B$.  Thus, the enhancement factor also suppresses the non-adiabatic contribution of the rotation on the nuclear spin, so for large magnetic fields, the range for detectable rotation rates is significantly enhanced.  In Fig.~\ref{fig:2}(c), we quantify the adiabatic range by $\xi^2 \sim 1\times10^{-5}$ from choosing the regime where $\langle \Delta S \rangle < 0.01$, thus allowing us to obtain the $\Omega_{\text{max}}$ in main text Fig.~3(a)i.

For fast rotations, we require the nuclear spin to remain in its initial eigenstate and decoupled from the electron spin rotation.  In this regime, as $\Omega \gg \gamma_e B$, the effective nuclear Hamiltonian becomes time-independent $H(t) \approx -\Omega I_y$, and $\mathcal{M}(t) \ll 1$, we can use similar analysis to characterize a minimum rotation rate for fast-rotation sensing $\Omega_{\text{min}}$ shown in main text Fig.~3(a)ii. In the main text, we refer to $\mathcal{M}(t)$ as the eigenstate deviation. 

\section{Simulation Details}
To simulate the dynamics of the system we can either work in the lab frame where the diamond, 
and thus the zero-field splitting and hyperfine interaction tensors, rotate; 
or work in the diamond frame, where the NV sees a rotating external magnetic field, as well as an effective field along the rotation axis due to the non-intertial frame. 

In the following, we discuss the basic principles of both methods and use them to simulate the system dynamics and study the spin relaxation effect.
\subsection{Lab Frame}
The dynamics of a quantum system is described by the Hamiltonian of the system. The ground state Hamiltonian of the $^{15}$NV center can be written as
\begin{equation}
\label{eq_supp:H}
    H = \Vec{S}\cdot \bold{D}\cdot\Vec{S} + \gamma_e \Vec{B} \cdot \Vec{S} + \gamma_n \Vec{B} \cdot \Vec{I} + \Vec{S} \cdot \bold{A} \cdot \Vec{I}
\end{equation}
where $\gamma_e=2 \pi \times 2.8024 $~MHz/G and $\gamma_n=2\pi \times 0.4316$~kHz/G are the gyromagnetic ratios of the electronic and nuclear spin. When the frame is chosen such that the $z$ axis is along the N-V orientation, the effective zero-field splitting (ZFS) tensor $\bold{D}$ is diagonal with the only nonzero diagonal term $D=2\pi \times 2.87$~GHz along the $z$ direction,  and the hyperfine tensor $\bold{A}$ includes only diagonal terms as well with $A_{zz}=2\pi \times 3.03$~MHz and $A_{xx} = A_{yy} = A_{\perp}=2\pi \times 3.65$~MHz.

The rotation of the diamond crystal can be described by the Euler angle $(\alpha,\beta,\gamma)$, represented by a rotation matrix $R(\alpha,\beta,\gamma) =R_z(\gamma)R_y(\beta)R_z(\alpha)$ 
with
\begin{equation}
   R_z(\alpha)= \left(
\begin{array}{ccc}
 \cos (\alpha) & -\sin (\alpha) & 0 \\
 \sin (\alpha) & \cos (\alpha) & 0 \\
 0 & 0 & 1 \\
\end{array}
\right),\quad
   R_y(\beta)=\left(
\begin{array}{ccc}
 \cos (\beta) & 0 & \sin (\beta) \\
 0 & 1 & 0 \\
 -\sin (\beta) & 0 & \cos (\beta) \\
\end{array}
\right)
\end{equation}
such that
\begin{equation}
    R(\alpha,\beta,\gamma) = \left(
\begin{array}{ccc}
 \cos (\alpha ) \cos (\beta ) \cos (\gamma )-\sin (\alpha ) \sin (\gamma ) & -\sin (\alpha ) \cos (\beta ) \cos (\gamma )-\cos (\alpha ) \sin (\gamma ) & \sin (\beta ) \cos (\gamma ) \\
 \cos (\alpha ) \cos (\beta ) \sin (\gamma )+\sin (\alpha ) \cos (\gamma ) & \cos (\alpha ) \cos (\gamma )-\sin (\alpha ) \cos (\beta ) \sin (\gamma ) & \sin (\beta ) \sin (\gamma ) \\
 -\cos (\alpha ) \sin (\beta ) & \sin (\alpha ) \sin (\beta ) & \cos (\beta ) \\
\end{array}
\right).
\end{equation}
Both of the ZFS and hyperfine interaction in the NV Hamiltonian are represented by 3-by-3 tensors, and as we introduced above both of them are diagonal in the reference frame choosing N-V as the $z$ axis. To clarify the difference between different frames, here we denote terms in the N-V frame with ``$\prime$" such that
\begin{equation}
    \bold{D^\prime}=\left(
\begin{array}{ccc}
 0 &0&0 \\
 0&0&0 \\
 0&0&D
\end{array}
\right), \quad \bold{A^\prime}=\left(
\begin{array}{ccc}
 A_{xx} &0&0 \\
 0&A_{yy}&0 \\
 0&0&A_{zz}
\end{array}
\right).
\end{equation}
The ZFS and hyperfine terms of the Hamiltonian in the NV frame can be written as
\begin{equation}
    (S_x^\prime,S_y^\prime,S_z^\prime)\cdot \bold{D^\prime}\cdot \left(
\begin{array}{c}
 S_x^\prime\\S_y^\prime\\S_z^\prime
\end{array}
\right)^\prime, \quad 
(S_x^\prime,S_y^\prime,S_z^\prime)\cdot \bold{A^\prime}\cdot \left(
\begin{array}{c}
 I_x^\prime\\I_y^\prime\\I_z^\prime
\end{array}
\right).
\end{equation}
Using $(S_x^\prime,S_y^\prime,S_z^\prime)^T = R^T(S_x,S_y,S_z)$, we can get the ZFS and hyperfine tensors in the lab frame
\begin{equation}
\label{eq_supp:DA_Lab}
    \bold{D}=R\cdot \bold{D}^\prime \cdot R^T, \quad \bold{A}=R\cdot \bold{A}^\prime \cdot R^T.
\end{equation}
The full Hamiltonian is then obtained by plugging these tensors into Eq.~\eqref{eq_supp:H}.
\label{sec: lab details}
\subsection{NV Frame}
Instead of working in the lab frame and calculating the co-rotating ZFS and hyperfine tensors, we can work in the NV frame to avoid complicated tensor rotations. Instead, in this case, the external static magnetic field has a rotation described by $R^T$ which is opposite to the diamond rotation, and the rotating frame transformation leads to additional terms in the Hamiltonian.

To simplify the analysis, here we consider a one-axis rotation ${\omega}$ along $y$. In the NV frame, the system Hamiltonian can be written as
\begin{equation}
    H = DS_z^2 + A_{zz}S_zI_z+A_\perp(S_xI_x+S_yI_y) + \gamma_e B\left[\cos(\omega t)S_z-\sin(\omega t)S_x\right] + \gamma_n B\left[\cos(\omega t)I_z-\sin(\omega t)I_x\right] -\omega S_y-\omega I_y.
\end{equation}
The simulation in the NV frame should give the same results as the simulation in the lab frame. In Fig.~\ref{fig_supp:gyro} we show a simulation of the nuclear spin evolution with the same parameters as in the main text, which shows consistent behavior.

\begin{figure}[h]
\includegraphics[width=0.6\textwidth]{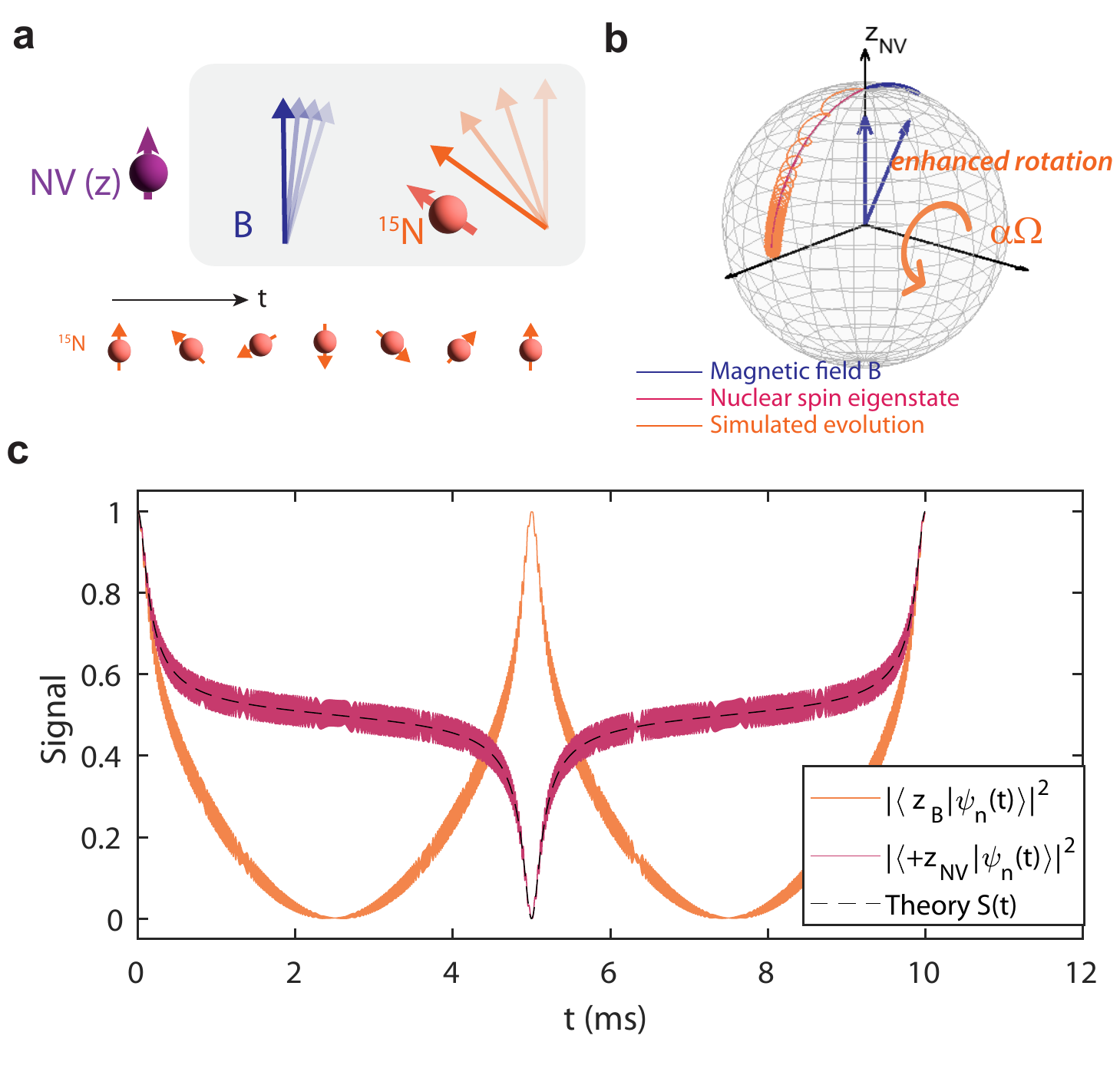}
\caption{\label{fig_supp:gyro} \textbf{Gyroscope with hyperfine-enhanced sensitivity simulated in the NV frame.} a. Effective dynamics of the electro-nuclear system for the hyperfine-enhanced gyroscope in the NV frame.  In the frame of the NV, the magnetic field counter rotates to the rotation, while the nuclear spin's rotation is enhanced from the magnetic field rotation. b. Simulation of the nuclear spin state evolution on the Bloch sphere. The diamond is rotated about $y$ axis with a rate $\omega=2\pi \times 0.1$~kHz, and the magnetic field is set to $B=50$~G. c. The overlap of the final  nuclear spin state with either its initial state or the state along the $B$ field direction.  }
\end{figure}
\label{sec: nv frame}
\subsection{Master Equation for Relaxation Time}
 While the coherence evolution of a closed system can be calculated using the Schr\"{o}dinger's equation, the evolution of an open system (e.g., depolarization and/or dephasing processes that  exist due to the system-bath coupling) can be derived using the Lindblad equation under the assumption of Markov bath,
\begin{equation}
\label{eq_supp:Lindblad}
    \frac{\partial}{\partial t}\rho(t)=-i[H(t),\rho(t)]+\sum_{k=1}^M\left(L_k\rho(t)L_k^\dagger-\frac{1}{2}L_k^\dagger L_k\rho(t)-\frac12\rho(t)L_k^\dagger L_k\right),
\end{equation}
where $L_k$ are Lindblad operators. To describe the spin relaxation dynamics in our system, the Lindblad operators are $L_k=\sqrt\Gamma\ket{m_s}\bra{m_s^\prime}$ where $m_s, m_s^\prime=\{-1,0,+1\}$ and the jump rate $\Gamma=1/(3T_{1e})$~\cite{chen_protecting_2018}.

In a practical simulation, the differential equation in Eq.~\eqref{eq_supp:Lindblad} is converted to an integral equation and is solved by calculating $\rho(t+\delta t)$ from $\rho(t)$ stepwise. The simulation precision depends on the step size $\delta t$.  

Due to the large order of magnitude difference between the system energy scale ($\sim$GHz) and the rotation rate ($<$kHz) which is of interest in our study, numerically calculating the evolution using the Lindblad equation in Eq.~\eqref{eq_supp:Lindblad} is challenging. To improve the simulation efficiency, we modify the Lindblad equation to eliminate the second approximation. The detailed derivation is as follows.

\subsubsection{Derivation}
The system evolution of an open quantum system can be described by a superoperator map. According to the Kraus Representation Theorem, a concrete expression of such a map can be written in the form of 
\begin{equation}
    S[\rho]=\sum_{k=1}^{K}M_k\rho M_k^\dagger\text{, with } \sum_{k=1}^{K}M_k^\dagger M_k=I
\end{equation}  
where $M_k$ are Kraus operators.

We can further write the evolution of $\rho$ from $t$ to $t+\delta t$ as 
\begin{equation}
    \rho(t+\delta t)=\sum_k M_k(\delta t) \rho(t) M_k^\dagger(\delta t).
\end{equation}
To separately account for contributions of the coherent evolution under the system Hamiltonian and the dephasing/depolarization process due to system-bath coupling, we separate the set of $M_k$ to $M_0$ and $M_k$ ($k>1$). The coherent part is included in $M_0=e^{-iH(t)\delta t}+\delta t K + O(\delta t^2)$ with $K$ hermitian, and the system-bath coupling effects are included in $M_k=\sqrt{\delta t}L_k +O(\delta t)$.
To satisfy the Kraus sum normalization condition $\sum M_k^\dagger M_k=I$, yielding
\begin{equation}
    M_0^\dagger M_0+\sum_{k>0} M_k^\dagger M_k=I +\delta t (e^{iH(t)\delta t}K+K^\dagger e^{iH(t)\delta t})+\delta t L_k^\dagger L_k+O(\delta t^2)=I +\delta t (K+K^\dagger )+\delta t L_k^\dagger L_k+O(\delta t^2),
\end{equation}
we obtain $K=-\frac{1}{2}\sum_{k>0}L_k^\dagger L_k$.
The evolution of $\rho$ is then calculated by 
\begin{equation}
\label{eq_supp:step}
    \rho(t+\delta t)=M_0\rho(t)M_0^\dagger +\sum_{k>0}M_k\rho(t)M_k^\dagger=e^{-iH(t)\delta t}\rho(t)e^{iH(t)\delta t}+\delta t\sum_{k>0}\left(L_k\rho(t)L_k^\dagger-\frac{1}{2}L_k^\dagger L_k \rho(t)-\frac{1}{2}\rho(t)L_k^\dagger L_k\right).
\end{equation}
When the jump rate is set to zero, the evolution in Eq.~\eqref{eq_supp:step} comes back to the form when considering only the coherent evolution under the system Hamiltonian, only requiring the step size $\delta t$ smaller than the characteristic time scale $T$ of the change in $H(t)$. And the second term linear in $\delta t$ requires that $\delta t$ should be much smaller than the characteristic jump time $1/\Gamma$.


\subsubsection{Simulation Results}
\label{supp_sec:sim_results}
Simulations were performed 
for varying values of $T_{1e}$ to compute the system coherence time.  Simulations using a modified Lindblad equation in Eq.~\eqref{eq_supp:step} demonstrated a $\sim 85 \%$ improvement in time efficiency per repetition.  In the adiabatic regime (at $B = 50$G and $\Omega = 0.1$kHz), we find the decay time of the signal yields consistently $\tau \approx 1.5 T_{1e}$ Fig.~\ref{fig_supp:ratio}(b), consistent with predictions from the spin fluctuator model~\cite{chen_protecting_2018}.   Nevertheless, in the inertial regime under fast rotation in Fig.~\ref{fig_supp:ratio}(a), the ratio $\tau/T_{1e}$ is a bit more unclear due to dynamical decoupling effects from the rapid rotation of the electron spin; our results yield that decreases with $T_{1e}$, as well as the rotation rate $\Omega$.  
\begin{figure}[h]
\includegraphics[width=0.80\textwidth]{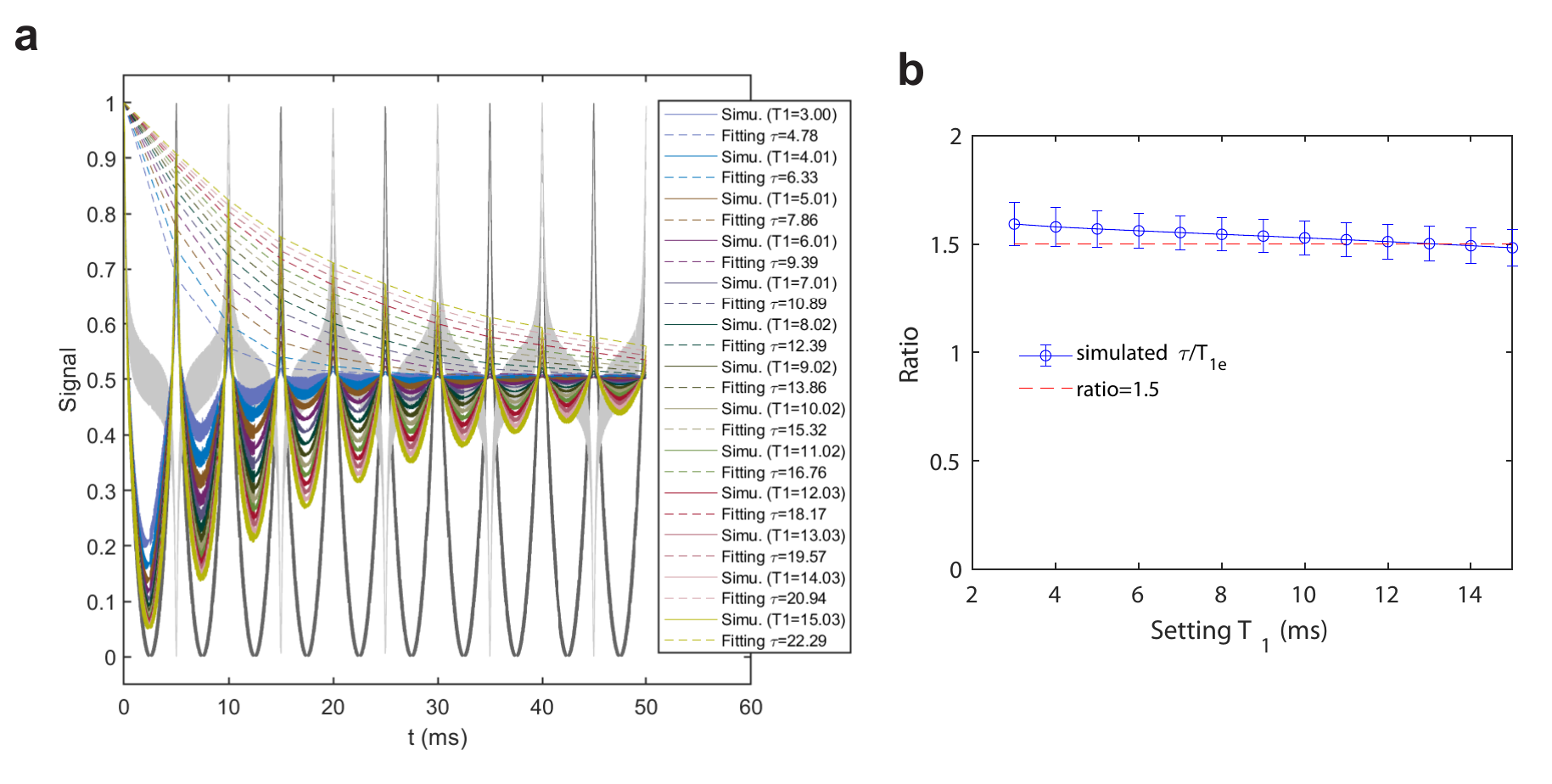}
\caption{\label{fig_supp:ratio} \textbf{Hyperfine-enhanced regime coherence time simulation} (a) Decay of the gyroscope signal and fitted coherence times (following envelop fit $S(t) = c_0 + c_1e^{-t/\tau}$) at $B = 50$G and $\Omega = (2\pi) 0.1$kHz for various $T_{1e}$ values. (b) Simulated ratios $\tau/T_{1e}$ from (a) as a function $T_{1e}$.}
\end{figure}

\subsection{Nuclear Spin Dephasing}
\label{sec: nuclear spin }
We also consider the effect of inhomogenous dephasing due to static thermal noise in the adiabatic regime.  We compute the system coherence time under a non-Markovian process where the we impose static local field fluctuations across the ensemble.  In particular, we simulate the evolution of the Hamiltonian averaged over $N$ spins where for each spin $i$:
\begin{equation}
    H_i = H_0 + \delta B_i \cdot \Vec{S}+ \delta B_i \cdot \Vec{I}
\end{equation}
where $B_i$ is sampled from a Gaussian distribution with mean $\langle \delta B \rangle = 0$ and we can describe the ensemble density matrix 
\begin{equation}
    \rho(t) = \int e^{-i H_i t}\rho(0) e^{i H_i t} P(\delta B) d\delta B
\end{equation}
where $P(\delta B) = \frac{e^{-(\delta B)^2/2(\Delta_B)^2}}{\sqrt{2 \pi}\Delta_B}$ where $\Delta_B = \sqrt{\langle \delta B^2 \rangle}$ is the strength (variance) of the magnetic field fluctuations.  For a typical NV ensemble sample, $\Delta_B \sim 0.1 - 0.5$G.  

During the system evolution under a bias magnetic field $B$, the nuclear spin precesses with an effective Larmor frequency $\omega_{eff} = \gamma_n B_{eff}$ where $B_{eff} = \sqrt{B_z^2 + \alpha^2 (B_x^2 + B_y^2)}$ due to the transverse hyperfine enhancement.  In the NV frame, the external magnetic field rotates counter-clockwise, and the nuclear spin precesses about the axis defined by $B_x  = -B\sin(\Omega t) \hat{x}$, $B_z  = B\cos(\Omega t)\hat{z}$.   In cases where  $B_{eff} \gg \Delta_B$, because we assume that the static noise does not rotate with the diamond (and thus rotates with the bias magnetic field in the NV frame), the large energy gap of the nuclear spin suppresses the influence of the transverse noise {$\delta B_x$ and $\delta B_y$}.  Nevertheless, the $\delta B_z$ noise causes a pure dephasing process, and the ensemble of spins precess at different frequencies, resulting in a decay in the transverse spin components ($T_2^*$ process).  Under a Gaussian distribution for $\delta B$, we can characterize the decay $T_2^* = \sqrt{2}/\Delta_B$ shown in Fig. \ref{fig:4}b.  

Thus during the gyroscope protocol, longitudinal noise results in a pure dephasing process for the nuclear spin, which reduces the effect of the ensemble nuclear precession in the measured signal.  In particular, as seen in Fig.~\ref{fig_supp:gyro}, the simulated dynamics of the system results in a signal with oscillations largest at $\theta = n\pi/2$.  In the presence of noise, these oscillations are suppressed by the $T_2^*$ dephasing process.  Thus, even though the transverse Bloch vector components of the nuclear spin decays during the sensing protocol, a pure dephasing process does not result in a decay of the overall measured signal that is used to extract the rotation rate and our protocol is robust against static magnetic field fluctuations, in contrast to previous NV-gyro protocols. 

{In our simulations in main text Fig.~3(b), we assume that $\delta B_x, \delta B_y = 0$ for an ensemble of $N = 1000$ spins, and we compute $T_2^*$ by measuring the decay of the oscillations in the signal at $\theta = n\pi/2$, where the nuclear spin experiences the same effective magnetic field strength. Varying $\Delta_{B_z}$, we confirm that fluctuations in $B_z$ results in a $T_2^*$ decay process.   Despite the enhancement factor's amplification of the transverse noise, because the effective magnetic field is also amplified by the same magnitude, as long as $B_{eff} \gg \delta B_{x,y}$, the dephasing process will be dominated by the longitudinal noise.  Considering only $\delta B_{x,y}$ noise, we see that the dephasing of the signal is much weaker.  Furthermore, in cases where the static fluctuations rotate with the diamond (thus remaining static in the NV frame), the components of $\delta B$ change in strength with the rotation, but ultimately, a pure dephasing process dominated by the longitudinal noise persists.  }

\end{widetext}

\end{document}